\documentclass[11pt,letterpaper]{article}
\pdfoutput=1

\usepackage{jcappub}
\usepackage{deluxetable}

\newcommand{\diff}{\mathrm{d}}
\newcommand{\geff}{G_{\rm eff}}
\newcommand{\rZ}{\mathcal{R}_{\rm SI\nu}}
\newcommand{\Bfac}{\mathcal{B}_{{\rm SI}\nu}}

\title{A tale of two modes: Neutrino free-streaming in the early universe}
\author{Lachlan Lancaster$^{1,2}$, Francis-Yan Cyr-Racine$^3$, Lloyd Knox$^4$, and Zhen Pan$^4$}
\affiliation{$^1$ McWilliams Center for Cosmology, Department of Physics, Carnegie Mellon University, 5000 Forbes Ave., Pittsburgh, PA 15213, USA}
\affiliation{$^2$ Institute of Astronomy, University of Cambridge, Madingley Road, Cambridge, CB3 0HA, United Kingdom}
\affiliation{$^3$ Department of Physics, Harvard University, 17 Oxford St., Cambridge, MA 02138, USA}
\affiliation{$^4$ Physics Department, University of California, Davis, CA 95616, USA}
\emailAdd{lachlanl@princeton.edu}
\emailAdd{fcyrraci@physics.harvard.edu}
\emailAdd{lknox@ucdavis.edu}
\emailAdd{zhpan@ucdavis.edu}

\abstract{We present updated constraints on the free-streaming nature of cosmological neutrinos from cosmic microwave background (CMB) temperature and polarization power spectra, baryonic acoustic oscillation data, and distance ladder measurements of the Hubble constant.  Specifically, we consider a Fermi-like four-fermion interaction between massless neutrinos, characterized by an effective coupling constant $\geff$, and resulting in a neutrino opacity $\dot{\tau}_\nu\propto \geff^2 T_\nu^5$.  Using a conservative flat prior on the parameter $\log_{10}\left( \geff {\rm MeV}^2\right)$, we find a bimodal posterior distribution with two clearly separated regions of high probability. The first of these modes is consistent with the standard $\Lambda$CDM cosmology and corresponds to neutrinos decoupling at redshift $z_{\nu,{\rm dec}} > 1.3\times10^5$, that is before the Fourier modes probed by the CMB damping tail enter the causal horizon. The other mode of the posterior, dubbed the ``interacting neutrino mode'', corresponds to neutrino decoupling occurring within a narrow redshift window centered around $z_{\nu,{\rm dec}}\sim8300$. This mode is characterized by a high value of the effective neutrino coupling constant, $\log_{10}\left( \geff {\rm MeV}^2\right) = -1.72 \pm 0.10$ (68\% C.L.), together with a lower value of the scalar spectral index and amplitude of fluctuations, and a higher value of the Hubble parameter.  Using both a maximum likelihood analysis and the ratio of the two mode's Bayesian evidence, we find the interacting neutrino mode to be statistically disfavored compared to the standard $\Lambda$CDM cosmology, and determine this result to be largely driven by the low-$l$ CMB temperature data. Interestingly, the addition of CMB polarization and direct Hubble constant measurements significantly raises the statistical significance of this secondary mode, indicating that new physics in the neutrino sector could help explain the difference between local measurements of $H_0$, and those inferred from CMB data. A robust consequence of our results is that neutrinos must be free streaming long before the epoch of matter-radiation equality in order to fit current cosmological data.
}


\begin{document}
\maketitle

\section{Introduction}
Due to their extremely weak interactions with other known particles of the Standard Model (SM),
deepening our understanding of neutrinos has become a game of large-scale experiments and statistics.
The cosmic microwave background (CMB) and other probes of the large-scale structure of the Universe have
proven themselves to be excellent probes of certain aspects of neutrino physics.
This is best exemplified by the limits that have been placed on the sum of neutrino masses
$\sum m_{\nu} \lesssim 0.15 $ eV \cite{Vagnozzi:2017ovm,DiValentino:2016ikp,Giusarma:2016phn,DiValentino:2015wba,DiValentino:2015sam,dePutter:2012sh,Giusarma:2012ph}, or by the constraints on the effective number of neutrino species $N_{\rm eff} = 3.04\pm0.18$ \cite{planck15}.

Within the SM, neutrinos interact exclusively through the electroweak force mediated by
the $W$ and $Z$ vector bosons \cite{Agashe:2014kda}. These interactions, due to their weak nature,
fall out of equilibrium early on in the history of the Universe when the SM plasma temperature was
$T \approx 1.5$ ${\rm MeV}$ \cite{Cyburt:2015mya}. After decoupling from other SM particles,
neutrinos are thought to free stream throughout the Universe, only interacting with other
species through their gravitational interactions. Within the standard cosmological paradigm,
free-streaming neutrinos are the primary source of anisotropic stress in the early Universe.
This causes the neutrino background to have a significant impact on the growth of cosmological
perturbations in the radiation dominated universe. Most notably, a nonvanishing anisotropic stress
results in an offset between the two Bardeen potentials in the conformal Newtonian gauge
(see e.g.~Ref.~\cite{Ma:1995ey}), which in turn affects the evolution of CMB fluctuations
leading to measurable amplitude shifts in the CMB temperature and polarization spectra
\citep{1972ApJ...176..323S,1973ApJ...180....1P,Chacko:2003dt,Bashinsky:2003tk,Friedland:2007vv,Hou:2011ec}. The supersonic (relative to the photon-baryon plasma) perturbations of free-streaming neutrinos additionally cause a phase shift in the oscillations of the power spectra which would not otherwise be
observed \citep{Bashinsky:2007yc,Bashinsky:2003tk, Follin:2015hya, Baumann15, Pan16,2017arXiv170310732B}. Moreover, the absence of neutrino clustering on scales  entering the causal horizon before neutrinos become nonrelativistic leads to a suppression of the matter power spectrum which could in principle be observed
with large-scale structure and CMB lensing observations \cite{Kaplinghat:2003bh}.

So far, the standard evolution of cosmological neutrinos outlined above appears in good agreement with observations 
\cite{Trotta:2004ty,Basboll:2008fx,Archidiacono:2013fha,planck15}. Still, the existence of neutrino mass 
\cite{Gribov:1968kq,Pontecorvo:1967fh,Ahmad:2001an} implies that some sort of physics beyond the SM must 
couple to neutrinos. Minimal modifications to the SM have been proposed  (see e.g.~refs.~\cite{GellMann:1980vs,Yanagida:1980xy,PhysRevLett.44.912,PhysRevD.22.2227}) to accommodate neutrino masses in ways that 
leave the standard evolution of cosmological neutrinos essentially unchanged. However, it is possible that the 
physics giving rise to neutrino mass does change the cosmological evolution of the neutrino bath, leading to 
detectable effects on the CMB and large-scale structures. For instance, neutrino interactions mediated by a not-yet-
observed particle 
\cite{BialynickaBirula:1964zz,Bardin:1970wq,Gelmini:1980re,Chikashige:1980qk,Barger:1981vd,Raffelt:1987ah,Kolb:1987qy,Konoplich:1988mj,Berkov:1988sd,Belotsky:2001fb,Chacko:2003dt,Beacom:2004yd, Hannestad:2004qu,Chacko:2004cz,Hannestad:2005ex, Bell:2005dr,Sawyer:2006ju,Cirelli:2006kt,Mangano:2006mp,Friedland:2007vv,Hooper:2007jr,Basboll:2008fx,Serra:2009uu,Aarssen:2012fx,Jeong:2013eza,Archidiacono:2013dua,Laha:2013xua,FYCR,Archidiacono:2014nda,Ioka:2014kca,Ng:2014pca,Cherry:2014xra,oldengott15,Archidiacono:2015oma,Cherry:2016jol,Archidiacono:2016kkh,Dvali:2016uhn,Capozzi:2017auw}
 could dramatically alter the free-streaming property of neutrinos on scales probed by the CMB. The key questions to ask then are whether current and upcoming data can pinpoint this new physics, and whether the latter can bias current and future constraints on the sum of neutrino masses.

Several works 
\cite{Chacko:2003dt,Beacom:2004yd,Bell:2005dr,Cirelli:2006kt,FYCR,Archidiacono:2013dua,Forastieri:2015paa,Forastieri:2017oma}
 have shown that current cosmological data allow for some flexibility in the underlying theoretical assumption of neutrinos interacting solely via the weak interaction. Previous studies have usually taken a very phenomenological approach to constrain models of neutrino properties, relying either on the effective fluid parameters $c_{\rm eff}$ and $c_{\rm vis}$ (the rest-frame sound-speed and viscosity parameter, respectively) 
\cite{Trotta:2004ty,Melchiorri:2006xs,DeBernardis:2008ys,Smith:2011es,Archidiacono:2011gq,Archidiacono:2012gv,Gerbino:2013ova,melchiorri14,Sellentin:2014gaa,planck15,2015JCAP...03..036A}, or on an effective redshift $z_{\rm eff}$ 
\cite{Basboll:2008fx,Archidiacono:2013dua} meant to characterize the kinetic decoupling or re-coupling epoch of neutrinos, depending on the model considered. While the former approach based on the effective fluid parameters has shown consistency with the standard neutrino cosmology, it is in general difficult to interpret these results in terms of new interactions in the neutrino sector \cite{FYCR,oldengott15}. The analyses using an effective re-coupling or decoupling redshift can capture more general neutrino interaction scenarios, but still leave out the details of the neutrino visibility function which could contain important clues on the nature of the yet-unseen neutrino interaction.

In this work, we compute updated cosmological constraints on the free-streaming nature of neutrinos in the early Universe. We focus on models where a new Fermi-like four-fermion interaction leads to a neutrino self-interaction rate $\Gamma_\nu\propto G_{\rm eff}^2T_\nu^5$ which suppresses free streaming at early times (here $G_{\rm eff}$ is a Fermi-like constant controlling the strength of the new interaction). Such models could occur for instance if SM neutrinos couple to a new scalar or vector mediator with mass above $\sim1$ keV \cite{Bardin:1970wq,Kolb:1987qy,Mangano:2006mp,Hooper:2007jr,Aarssen:2012fx,Laha:2013xua}, or, perhaps more plausibly, if SM neutrinos mix with a sterile neutrino which is itself coupled to a new vector or scalar mediator \cite{Cherry:2014xra,Cherry:2016jol,Capozzi:2017auw}. We do not consider here models where neutrinos (either active or sterile) couple to a massless or nearly massless particle such as a majoron \cite{Gelmini:1980re,Chikashige:1980qk,Kolb:1987qy,Raffelt:1987ah,Chacko:2003dt,Beacom:2004yd,Hannestad:2004qu,Hannestad:2005ex,Friedland:2007vv,Archidiacono:2015oma,Archidiacono:2016kkh} since these typically lead to a neutrino self-interaction rate scaling as $\Gamma_\nu\propto T_\nu$ at energies relevant to the CMB. For simplicity, we also focus exclusively on massless neutrinos since the presence of neutrino mass significantly complicates the computation of the self-interaction collision integral. In an upcoming work \cite{Kreisch_in_prep}, we show that the results presented here are robust to the inclusion of neutrino masses. 

Cosmological constraints on neutrino interaction models with $\Gamma_\nu\propto G_{\rm eff}^2T_\nu^5$ were placed in refs.~\cite{FYCR,Archidiacono:2013dua}.  There, it was found that the posterior distribution for the interaction strength controlling neutrino self interaction was bimodal. While one of the modes shows broad consistency with the standard evolution of cosmological neutrinos, the second mode of the posterior corresponds to a radically different evolution of cosmological neutrinos in which neutrinos are strongly self interacting until close to the epoch of matter-radiation equality.  Assessing the statistical significance of this somewhat surprising ``interacting neutrino mode'' is complicated by the fact that it depends on the choice of the prior probability assigned to each mode \cite{FYCR}. Here, we adopt a conservative choice of prior and use the latest CMB, baryon acoustic oscillation (BAO), and Hubble parameter data to test for the presence of new physics in the neutrino sector.

Of course, CMB and other cosmological data are not the only probe of neutrino physics beyond the SM. A summary of constraints on ``secret'' neutrino interaction has been presented in ref.~\cite{Ng:2014pca}. For a fairly low mass mediator ($\lesssim 10$ MeV), SN 1987A \cite{Kolb:1987qy}, Big Bang nucleosynthesis \cite{Ahlgren:2013wba}, the CMB, and the detection of PeV neutrinos at IceCube \cite{Ng:2014pca,Cherry:2016jol} provide the strongest constraints, with the latter bound having the potential of being the most stringent. There are also flavor-dependent bounds based on measurements of lepton and meson decays \cite{Lessa:2007up}, although these could potentially be evaded by an appropriate choice of the flavor structure of the matrix coupling the new interaction to SM neutrinos. Other limits \cite{Bilenky:1992xn,Bardin:1970wq,Bilenky:1999dn} coming from Z-boson decay do not apply at the energy scale probed by the CMB. We note that elastic collisions caused by the new interaction do not change the time it takes for neutrinos to escape supernovae \cite{Dicus:1988jh}, although they could lead to new interesting phenomena (see e.g.~refs.~\cite{PhysRevLett.95.191302,PhysRevD.83.117702,PhysRevLett.96.211302,PhysRevD.42.293,Blennow:2008er}). Finally, supernova cooling puts bounds on the coupling of majorons to SM neutrinos \cite{Kachelriess:2000qc,Farzan:2002wx,Zhou:2011rc}, but the applicability of these to the models considered here is uncertain.

We begin in section \ref{sec:nunu} by discussing the generic properties of the neutrino interaction model that we consider here and by presenting the key equations describing the cosmological evolution of self-interacting neutrinos.
The data used in our analysis is introduced in section \ref{sec:data_anal}, while our results are given in section \ref{results}. We discuss the implications of our findings in section \ref{sec:discussion}. We finally perform a CMB stage-IV Fisher Matrix forecast in section \ref{forecasting} before concluding in section \ref{sec:conclusion}.

\section{Cosmological evolution with self-interacting neutrinos}
\label{sec:nunu}
\subsection{Interaction model}
We consider a phenomenological model in which SM neutrinos are coupled to a
new massive scalar field $\phi$ with mass $M_{\phi}$ via an interaction term of the form $\mathcal{L}_{\rm int} \sim y_\nu\phi\overline{\nu} \nu$, where  $y_{\nu}$ is a dimensionless coupling constant. As discussed in ref.~\cite{Ng:2014pca}, this interaction model should \emph{not} be taken at face value but instead serves as a rough guide to delimit the parameter space relevant to ``secret'' neutrino interactions. Within this phenomenological model, we focus on the low-energy effective theory where the neutrino temperature $T_{\nu}$ is significantly less than $M_\phi$, allowing us to integrate out the massive mediator and treat the interaction as a four-neutrino vertex with a dimensional coupling constant $G_{\nu}$ given by\footnote{We note that our definition of $G_\nu$ is slightly different than that used in ref.~\cite{Ng:2014pca}.}
\begin{equation}\label{Gnu}
G_{\nu} = \frac{y_{\nu}^2}{M_{\phi}^2}.
\end{equation}
As we focus on the regime where the neutrino temperature is significantly less than the mediator mass, we can safely ignore the presence of the mediator in the thermal bath. Of course, the annihilation of the $\phi$ particles when $T_\nu\sim M_\phi$ would generally heat the neutrino bath above its standard value, leading to a non-zero value of the number of extra relativistic degrees of freedom, $\Delta N_{\rm eff}$.  However, since the specifics of this heating strongly depend on the details of the neutrino interaction model, we do not include it here and thus fix $N_{\rm eff} = 3.046$ for the remainder of this paper. In an upcoming work \cite{Kreisch_in_prep}, we show that our results are robust to this particular choice.

For our Fermi-like phenomenological model, the thermally-averaged neutrino self-interac-tion cross section scales as $\langle \sigma_{\nu\nu} v\rangle \sim G_\nu^2 T_\nu^2$. Since the neutrino number density itself scales as $n_\nu\sim T_\nu^3$ after weak neutrino decoupling, we get that the neutrino self-interaction rate goes as $n_\nu \langle \sigma_{\nu\nu} v\rangle \sim G_\nu^2 T_\nu^5$. Any realistic model of neutrino interaction admitting an equivalent low-energy effective theory as our phenomenological model will lead to a similar interaction rate, up to a pre-factor depending on the details of the interaction. We thus define our neutrino self-interaction opacity as $\dot{\tau}_{\nu} \equiv -a \xi G_{\nu}^2 T_{\nu}^5$, where $a$ is the scale factor describing the expansion of the universe, and $\xi$ is a constant of order unity that is determined by the details of the neutrino interaction model \cite{FYCR}.  We note that the overhead dot denotes a derivative with respect to conformal time (which is why there is a scale factor appearing in our definition of opacity), and that the opacity is by convention negative since it is the derivative of the neutrino optical depth. Since the neutrino opacity only depends on the product of $\xi$ and $G_{\nu}$, we define a rescaled coupling constant
\begin{equation}
\label{Geff}
\geff \equiv \sqrt{\xi}G_{\nu},
\end{equation}
which allows us to write down the neutrino opacity as
\begin{equation}\label{nu_opacity}
\dot{\tau}_{\nu} =-a \geff^2 T_{\nu}^5.
\end{equation}
As we show below, the CMB is sensitive only to relatively large values of $\geff$, and it is thus reasonable to assume that $ \geff \gg G_{\rm F}$, where $G_{\rm F}\simeq1.166\times10^{-11}$ MeV$^{-2}$ is the standard Fermi constant. For the remainder of this work, we thus justifiably ignore the electroweak contribution to the neutrino opacity, as is traditionally done in standard CMB analyses.
\subsection{Cosmological perturbations}
We now turn our attention to the evolution of cosmological perturbations in the presence of self-interacting neutrinos. We adopt the notation of ref.~\cite{Ma:1995ey} in the synchronous gauge. The details of the derivation of the neutrino Boltzmann hierarchy in the presence of self-interaction are presented in ref.~\citep{oldengott15}. The presence of significant momentum transfer in a typical neutrino interaction renders the computation of the collision integrals rather tedious. However, since the neutrino opacity given in eq.~\eqref{nu_opacity} is a steep function of temperature, there is only a narrow window in time where the details of the neutrino interactions play a role. Indeed, at early times when $|\dot{\tau}_\nu| \gg \mathcal{H}$, the neutrinos form a tightly-coupled fluid such that only the two lowest moments (corresponding to energy density and heat flux) of the Boltzmann hierarchy are nonzero\footnote{Here, $\mathcal{H}$ is the conformal Hubble parameter.}. On the other hand, after neutrino decoupling when $|\dot{\tau}_\nu| \ll \mathcal{H}$, the collision term plays little role in the evolution of neutrino perturbations.  Only near the peak of the neutrino visibility function (see e.g.~figure \ref{fig:vis_funct} below) can the significant momentum transferred in a neutrino collision modify the neutrino phase-space distribution function. Even in this case, since we are only sensitive to the neutrino perturbations through their gravitational impact on CMB photons, it is unclear whether these distortions of the neutrino phase-space distribution would be observables. Moreover, the absence of an energy sink or source near the epoch of neutrino last-scattering all but guarantees that distortions of the neutrino distribution function would be small\footnote{We note for instance that for the SM neutrino decoupling near the epoch of nucleosynthesis, the bulk of the neutrino spectral distortions comes from the energy injected by $e^+e^-$ annihilation \cite{Mangano:2005cc}.}. 

While it would be interesting to compute whether the distortions to the neutrino phase-space distribution function can lead to observable effects in the CMB, we adopt here a simplified picture in which we assume that the neutrino distribution function exactly maintains its equilibrium thermal shape throughout the history of the Universe. This allows us to simplify the collision integrals, and write a momentum-integrated hierarchy of neutrino multipole moments.  This Boltzmann hierarchy takes the form
\begin{equation}\label{eq:l0}
\dot{\delta}_\nu = -\frac{4}{3}\theta_\nu - \frac{2}{3}\dot{h},
\end{equation}
\begin{equation}\label{eq:l1}
\dot{\theta}_\nu = k^2\left( \frac{1}{4}\delta_\nu - \frac{1}{2}\mathcal{F}_{\nu2}\right),
\end{equation}
\begin{equation}
\label{l2}
\dot{\mathcal{F}}_{\nu,2} = \frac{8}{15}\theta_\nu -\frac{3}{5} k\mathcal{F}_{\nu,3} +\frac{4}{15} \dot{h} +\frac{8}{5}\dot{\eta} + \alpha_2 \dot{\tau}_{\nu} \mathcal{F}_{\nu,2},
\end{equation}
\begin{equation}
\label{hiL}
\dot{\mathcal{F}}_{\nu,l} = \frac{k}{2l+1} \left[l \mathcal{F}_{\nu,(l-1)} - \left(l+1 \right)\mathcal{F}_{\nu, (l+1)} \right] + \alpha_l \dot{\tau}_{\nu} \mathcal{F}_{\nu, l},\ \ \ l \ge 3
\end{equation}
where $\delta_\nu$ is the neutrino energy density perturbation, $\theta_\nu$ is the divergence of the neutrino heat flux, $h$ and $\eta$ are the metric perturbation in the synchronous gauge \cite{Ma:1995ey}, $\mathcal{F}_{\nu,l}$ is the $l$th multipole moment of the neutrino Boltzmann hierarchy, $k$ is the magnitude of the comoving wave vector in Fourier space, and the $\alpha_l$ are dimensionless coefficients of order unity that depend on the specifics of the neutrino interaction model. Noting that a change to the value of $\alpha_2$ can be absorbed into the definition of $\geff$, and that the higher order coefficients $\alpha_{l\geq3}$ have a small impact on the CMB temperature and polarization spectra, we simply set $\alpha_l = 1$ for all $l$ in this work.

We solve equations (\ref{eq:l0}) through (\ref{hiL}) numerically along with perturbation equations for other standard cosmological components such as baryonic matter, photons, and cold dark matter using a version of the code \verb+CAMB+ modified for the purposes of this study \citep{camb12,camb00}. Due to the tight coupling of the self-interacting neutrinos whenever $|\dot{\tau}_\nu| \gg \mathcal{H}$, equations \eqref{l2} and \eqref{hiL} are difficult to solve numerically at early times. In this regime, we adopt a tight-coupling approximation scheme \cite{CyrRacine:2010bk} where we set $\mathcal{F}_{\nu,2} = 4 \left(2\theta_\nu+ \dot{h} + 6 \dot{\eta} \right)/\left(15 \alpha_2 \dot{\tau}_{\nu} \right)$ and $\mathcal{F}_{\nu,l}=0$ for $l \ge 3$. We switch off this approximation whenever the condition $|\dot{\tau}_\nu|/\mathcal{H}< 100$ is satisfied. We have checked that this switch occurs early enough as to not bias our results. Finally, we truncate our neutrino Boltzmann hierarchy using the procedure outlined in ref.~\cite{Ma:1995ey}.

\section{Data analysis methods}
\label{sec:data_anal}
As the analysis performed in ref.~\cite{FYCR} found evidence for a multimodal posterior distribution for the cosmological and neutrino parameters in the presence of self-interactions, we do not use the standard Markov Chain Monte Carlo method based on the Metropolis-Hastings (MH) algorithm to explore the parameter space. Indeed, the MH algorithm displays poor convergence properties for posteriors containing distinct islands of high probability separated by wide valleys of low probability, which is exactly the type of posterior that our self-interacting neutrino model produces, as we will discuss in section \ref{results}. To tackle this problem, we use the more robust nested-sampling algorithm \cite{skilling2006} as implemented in the software package MultiNest \citep{feroz08,feroz09,feroz13} to scan the parameter space and generate random samples of the posterior probability distribution function (PDF). As a by-product, nested sampling can also yield an estimate of the Bayesian evidence, which is useful for model comparison (see section \ref{sec:bayesian_evidence} below).

We employ the cosmological parameter estimation package \texttt{CosmoSIS} \citep{zuntz15} to wrap together  the MultiNest inference code, the \texttt{CAMB} Boltzmann solver, and the different data likelihoods used in this work (see section \ref{sec:dataset}). Our parameter space consists of the 6 standard $\Lambda$CDM cosmological parameters, together with the base-10 logarithm of the neutrino self-interaction strength, $\{ A_{\rm s}, n_{\rm s}, \tau_{\rm reio}, H_0,\Omega_{\rm b}h^2,\Omega_{\rm c}h^2, \log_{10} \left( \geff\ {\rm MeV^2}\right)\}$.  We use flat priors for the standard cosmological parameters as described in \cite{planck15Params}, as well as a flat prior on $\log_{10} \left( \geff\ {\rm MeV^2}\right) \in [-5.0,0.0] $. We justify this choice of prior by noting that models with $\geff < 10^{-4.5}$MeV$^{-2}$ have no impact on scales currently probed by the CMB, while models with $\geff > 10^{-1}$MeV$^{-2}$ have their neutrino decoupling epoch occurring in the matter-dominated era when neutrinos play a subdominant role in the evolution of CMB and matter fluctuations. We perform nested sampling with 2000 live points and terminate the chains when the accuracy reaches a tolerance of $10^{-1}$ in the logarithm of the Bayesian evidence. We have verified the stability of our results by both increasing the number of live points and lowering the lower bound of the prior on $\log_{10}\left( \geff {\rm MeV^2}\right)$ and subsequently checking that we obtain the same parameter constraints. 
\subsection{Data sets}\label{sec:dataset}
As our main dataset, we use CMB data from the Planck 2015 release \cite{planck15Params}. In particular, we include both the low-$\ell$ and high-$\ell$ temperature auto-correlation data, which we shall collectively refer to as the ``TT'' dataset\footnote{For clarity, we use the \texttt{commander\_rc2\_v1.1\_l2\_29\_B} and \texttt{plik\_lite\_v18\_TT} likelihoods at low-$\ell$ and high-$\ell$, respectively.}.  To reduce the computational burden, we use the ``lite'' version of the high-$\ell$ likelihood which internally marginalizes over the nuisance parameters of the data reduction \citep{planck15nuisance}. We also combine the CMB temperature data with the high-$\ell$ E-mode polarization auto-correlation and temperature cross-correlation data, a combination that we shall collectively refer to as ``TT + Pol''\footnote{In addition to the low-$\ell$ temperature likelihood, the exact likelihood used in this case is \texttt{plik\_lite\_v18\_TTTEEE}.}. 

We also use data from the Sloan Digital Sky Survey (SDSS-III) Baryon Oscillation Spectroscopic Survey (BOSS) data release 12 \cite{alam16}. In particular, we include their constraints on the volume averaged distance scale $D_{\rm V} (z)$, obtained from the measurement of the BAO feature at three different redshifts $z = 0.44$, $0.60$, and $0.73$. We shall refer to this dataset as ``BAO'' in the rest of the paper.  

Finally, we also use the direct measurement of the local Hubble Parameter from ref.~\cite{riess11}, $H_0 = 73.8\,\pm\,2.4$ km/s/Mpc, which is obtained from a traditional distance ladder approach using Cepheids and supernovae. We refer to this likelihood as ``$H_0$'' in what follows.

\section{Results}
\label{results}
\subsection{Posterior distributions and parameter limits}
\begin{figure}[t!]
\includegraphics[width=0.99\textwidth]{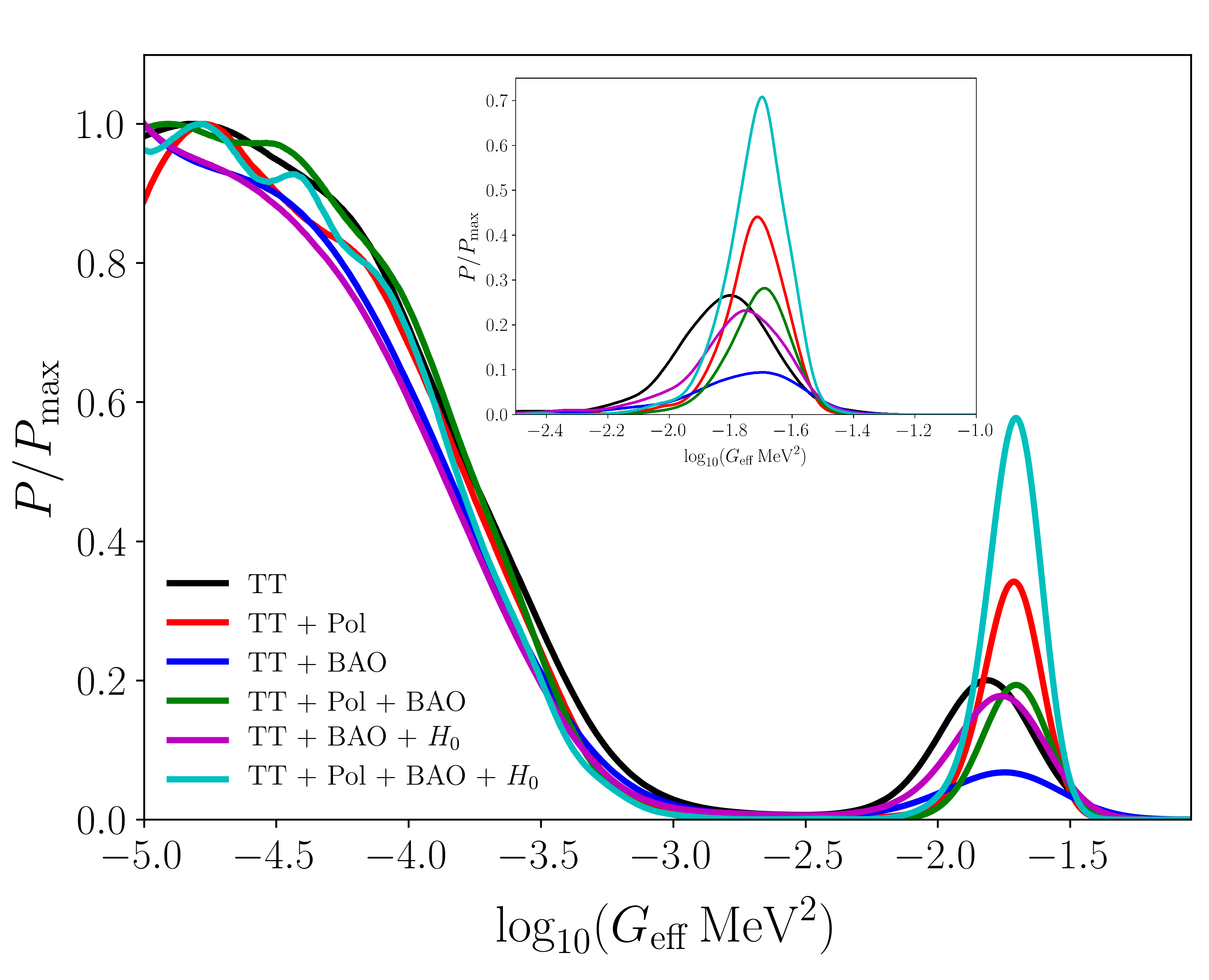}
\caption{Posterior probability distribution of the parameter $\log_{10}\left( \geff {\rm MeV}^2\right)$, marginalized over all other cosmological parameters, for the different combination of datasets used in this work. We assume a flat prior in $\log_{10}\left( \geff {\rm MeV}^2\right)$ with range $[-5,0]$. We note that the bimodality of the posterior implies that it is difficult to choose a unique smoothing kernel that is valid over the whole prior range. The smoothing kernel used here is a compromise between under smoothing the main mode and over smoothing the interacting neutrino mode. To illustrate this latter effect, we show in the inset the interacting neutrino mode plotted using a much narrower smoothing kernel appropriate to the width of this mode.}\label{fig:marggeff}
\end{figure}

We show in figure \ref{fig:marggeff} the marginalized posterior distribution of the parameter $\log_{10} \left( \geff {\rm MeV}^2\right)$ for the different combinations of likelihoods considered in this work. As found in ref.~\cite{FYCR}, we find that the posterior is bimodal, with a first mode characterized by small values of the effective neutrino coupling constant and consistent with the standard $\Lambda$CDM paradigm, and a second mode characterized by large values of the effective coupling constant and corresponding to a cosmology in which the onset of neutrino free-streaming is severely delayed compared to the standard scenario. We see that this ``interacting neutrino mode'' is present for every combination of data sets we used here, although its relative statistical weight compared to the standard $\Lambda$CDM mode varies from one data set to the next. We discuss the statistical significance of each mode further in section \ref{sec:bayesian_evidence} below. We note that since the two modes have very different widths, it is difficult to choose a single smoothing kernel for the whole range of the posterior shown, leading to an over-smoothing of the interacting neutrino mode. To illustrate this, we show in the inset of figure \ref{fig:marggeff} the posterior of the interacting neutrino mode plotted using a smoothing kernel appropriate for the width of that mode, leading to changes of the smoothed PDF at the $10$ to $20\%$ level. We however emphasize that the choice of smoothing kernel is only relevant when plotting the posterior, and the quantitative parameter limits presented in this work do not depend on the choice of kernel. 
\begin{deluxetable}{ccccc} 
\tabletypesize{\footnotesize} 
\tablecolumns{5} 
\tablewidth{0pt} 
\tablecaption{ Parameter constraints in the $\Lambda$CDM mode for  4 different data combinations. Unless otherwise noted, we display the $68\%$ confidence limits. \label{table:results1}} 
\tablehead{ 
\colhead{Parameter} & \colhead{TT} &  \colhead{TT + Pol  }& \colhead{TT + Pol + BAO} & \colhead{TT + Pol + BAO + $H_0$} } 
\startdata 
{$\Omega_{\rm b}h^2  $}  &  $0.02222\pm0.00027$ & $0.02223\pm0.00017$ & $0.02226\pm0.00014$ & $0.02231\pm0.00014$  \\
{$\Omega_{\rm c}h^2  $}  &  $0.1190\pm0.0026$ & $0.1193\pm0.0016$ & $0.1189\pm0.0011$ & $0.1183\pm0.0011$  \\
{$H_0$} [km/s/Mpc]  &  $68.1\pm1.2$ & $67.90\pm0.72$ & $68.11\pm0.50$ & $68.36\pm0.50 $  \\
{$\tau_{\rm reio} $}  &  $0.098\pm0.033$ & $0.095\pm0.024$ & $0.099\pm0.022$ & $0.104\pm0.022$  \\
{$n_{\rm s} $}  &  $0.9634\pm0.0082$ & $0.9620\pm0.0057$ & $0.9634\pm0.0047$ & $0.9650\pm0.0047$  \\$10^9A_{\rm s} $  &  $2.28\pm0.14$ & $2.27\pm0.10$ & $2.284\pm0.096$ & $2.304\pm0.098$  \\
$\log_{10}(\geff {\rm MeV}^2)$  &  $<-3.48$ (95\%) & $<-3.55$ (95\%) & $<-3.57$ (95\%) & $<-3.60$ (95\%)  
\enddata 
\vspace{-0.4cm} 
\end{deluxetable}
\begin{deluxetable}{ccccc} 
\tabletypesize{\footnotesize} 
\tablecolumns{5} 
\tablewidth{0pt} 
\tablecaption{ Parameter $68\%$ confidence limits within the interacting neutrino mode.  \label{table:results2}} 
\tablehead{ 
\colhead{Parameter} & \colhead{TT} &  \colhead{TT + Pol  }& \colhead{TT + Pol + BAO} & \colhead{TT + Pol + BAO + $H_0$} } 
\startdata 
{$\Omega_{\rm b}h^2  $}  &  $0.02256\pm0.00033$ & $0.02248\pm0.00017$ & $0.02240\pm0.00016$ & $0.02244\pm0.00016$  \\
{$\Omega_{\rm c}h^2  $}  &  $0.1177\pm0.0028$ & $0.1200\pm0.0017$ & $0.1210\pm0.0013$ & $0.1206\pm0.0012$  \\
{$H_0$} [km/s/Mpc]  &  $70.4\pm1.3$ & $69.59^{+0.74}_{-0.71}$ & $69.13\pm0.51$ & $69.33\pm0.52$  \\
{$\tau_{\rm reio} $}  &  $0.113\pm0.036$ & $0.103^{+0.022}_{-0.024}$ & $0.094^{+0.021}_{-0.023}$ & $0.098\pm0.021$  \\
{$n_{\rm s} $}  &  $0.9431^{+0.0091}_{-0.0084}$ & $0.9376\pm0.0054$ & $0.9344^{+0.0045}_{-0.0047}$ & $0.9359\pm0.0047$  \\
$10^9A_{\rm s} $  &  $2.21^{+0.15}_{-0.16}$ & $2.164^{+0.093}_{-0.10}$ & $2.131^{+0.087}_{-0.095}$ & $2.145\pm0.091$  \\
$\log_{10}(\geff {\rm MeV}^2)$  &  $-1.83\pm0.16$ & $-1.727^{+0.10}_{-0.092}$ & $-1.711^{+0.099}_{-0.11}$ & $-1.720^{+0.10}_{-0.094}$  
\enddata 
\vspace{-0.4cm} 
\end{deluxetable}

\begin{figure}[t!]
\center\includegraphics[width=0.8 \textwidth]{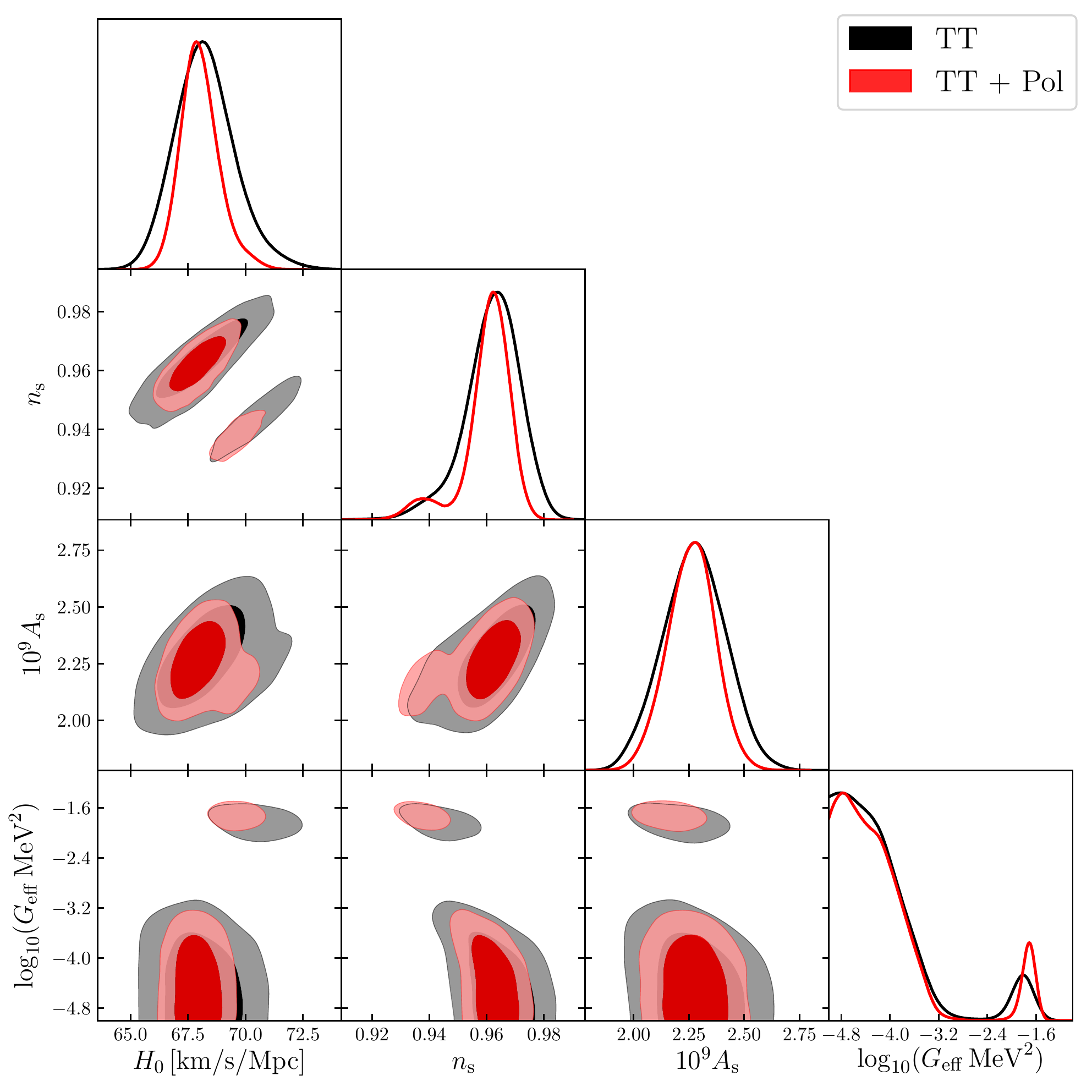}
\caption{Posterior distributions for the cosmological parameters most affected by neutrino self-interaction. We illustrate here the constraints obtained by using Planck temperature-only data (red), as well as the combination of Planck temperature and E-mode polarization data (black). The contours in the two-dimensional plots give the 68\% and 95\% confidence regions.}
\label{fig:triang_base}
\end{figure}

\begin{figure}[t!]
\center\includegraphics[width=0.8 \textwidth]{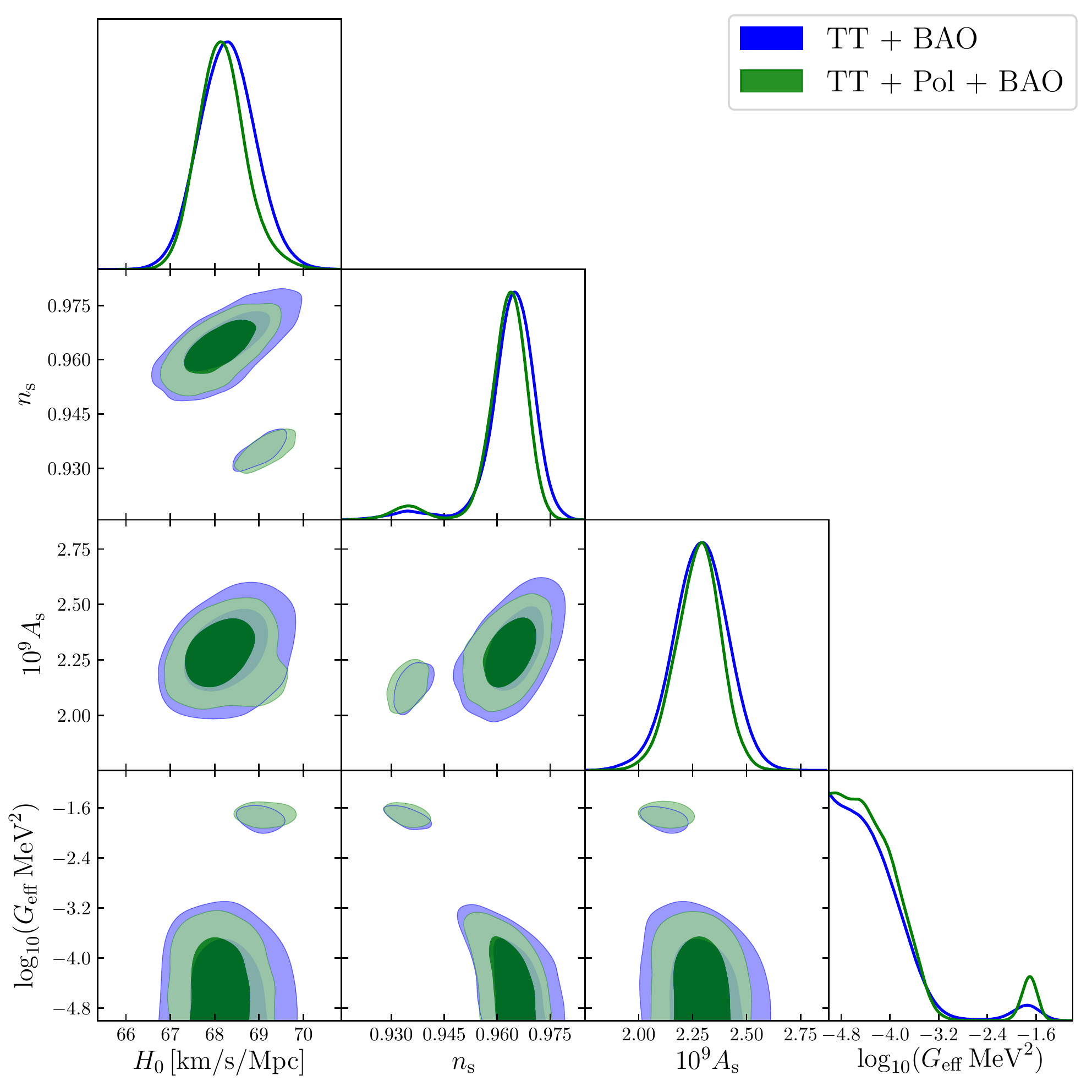}
\caption{Similar to the posterior distributions shown in figure \ref{fig:triang_base}, but adding the BAO measurements from the SDSS BOSS DR12 \cite{alam16}.}
\label{fig:triang_BAO}
\end{figure}

\begin{figure}[t!]
\center\includegraphics[width=0.8 \textwidth]{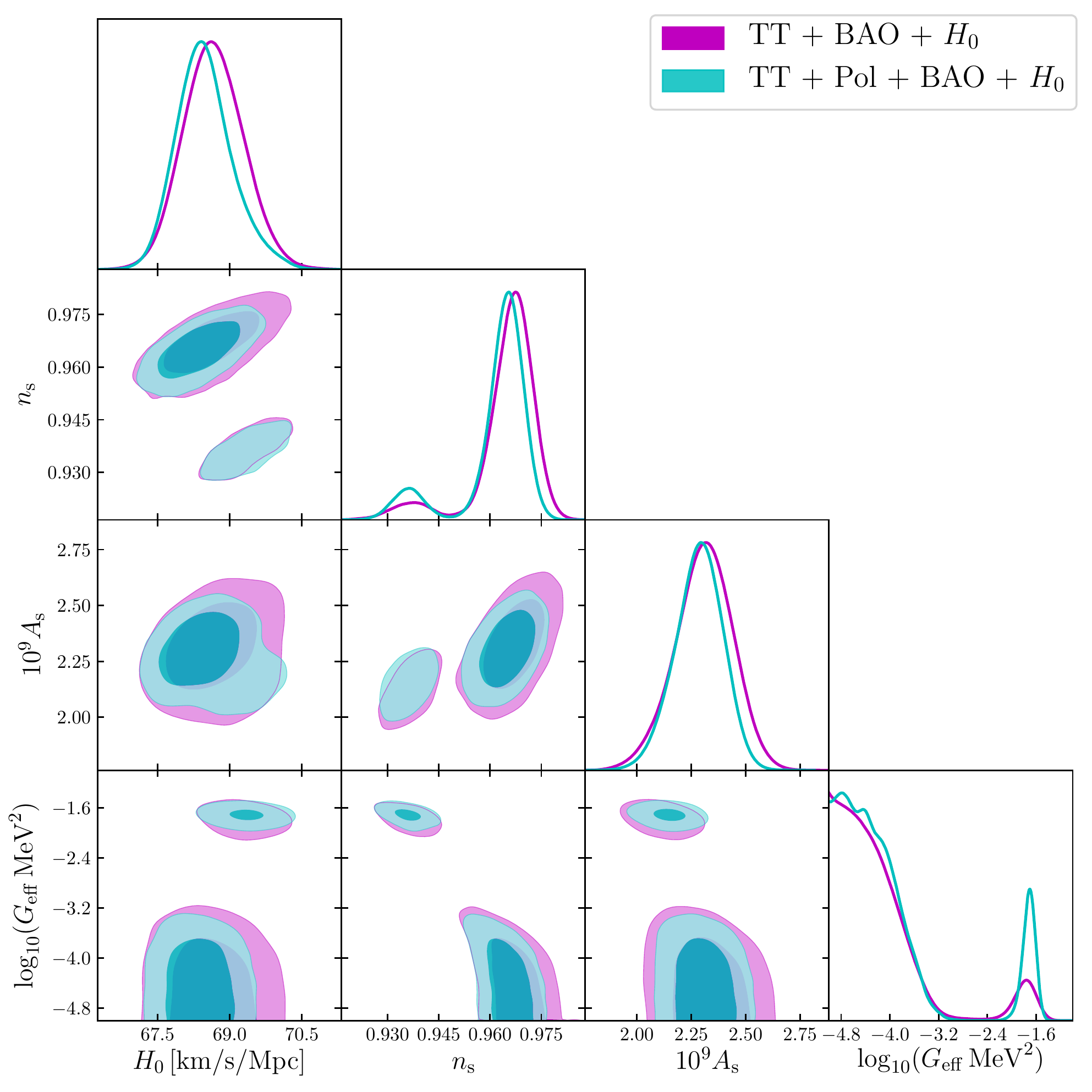}
\caption{Similar to the posterior distributions shown in figure \ref{fig:triang_base}, but adding the BAO measurements from the SDSS BOSS DR12 \cite{alam16} and the local measurement of the Hubble parameter from ref.~\cite{riess11}.}
\label{fig:triang_BAO_H0}
\end{figure}

In figures \ref{fig:triang_base}, \ref{fig:triang_BAO}, and \ref{fig:triang_BAO_H0}
we present one and two-dimensional marginalized PDFs for the four cosmological parameters
that are affected most by the introduction of neutrino self-interactions at early times: $H_0$, $A_{\rm s}$, $n_{\rm s}$, and $\log_{10} \left( \geff {\rm MeV}^2\right) $. In each of these plots we compare the effects of using Planck temperature and polarization data with simply using the temperature data (along with the other likelihoods that we list in the legend). The quantitative parameter confidence limits are given in table \ref{table:results1} for the $\Lambda$CDM mode, and in table \ref{table:results2} for the interacting neutrino mode. For the $\Lambda$CDM mode of the different posteriors we computed, the parameter constraints on $\Omega_{\rm b}h^2$, $\Omega_{\rm c}h^2$, $H_0$, and $n_{\rm s}$ are broadly consistent with those of the standard 6-parameter $\Lambda$CDM paradigm \cite{planck15}, with parameter shifts of less than one sigma. As visible in figures \ref{fig:triang_base}, \ref{fig:triang_BAO}, and \ref{fig:triang_BAO_H0}, larger values of $\log_{10}(\geff {\rm MeV}^2)$ within the $\Lambda$CDM mode are correlated with smaller values of the scalar spectral index. This reflects the impact of neutrino self-interaction on the amplitude of the CMB temperature spectrum: the absence of neutrino free-streaming damping \cite{Bashinsky:2003tk} is being compensated by a lower spectral index. However, the fact that the contours of the $\Lambda$CDM mode close at finite values of $\log_{10}(\geff {\rm MeV}^2)$ implies that it is not always possible to compensate for the absence of neutrino free-streaming by adjusting the primordial spectrum of fluctuations. We return to this point in section \ref{sec:discussion}.

For the $\Lambda$CDM mode, the values of the optical depth to CMB last scattering $\tau_{\rm reio}$ for all data set combinations are slightly larger than the values presented in ref.~\cite{planck15}, but still fully consistent given the error bars. These larger values of $\tau_{\rm reio}$ are compensated by somewhat higher amplitudes of the scalar fluctuations $A_{\rm s}$ (compared to ref.~\cite{planck15}) in order to obtain the correct amplitude of the CMB temperature spectrum, which itself depends on the product $A_{\rm s}e^{-2\tau_{\rm reio}}$. Within the $\Lambda$CDM mode, all dataset combinations considered here lead to an upper bound on the strength of the neutrino self-interaction approximately given by $\log_{10}(\geff {\rm MeV}^2) < -3.55$ (95\% C.L., see table \ref{table:results1} for the exact limits). This can be translated to a constraint on the neutrino decoupling redshift $z_{\nu,{\rm dec}} > 1.3\times10^5$. This limit is nearly identical to that presented in ref.~\cite{FYCR}, which is based on Planck 2013 CMB temperature data \cite{2014A&A...571A..16P}. It may appear surprising that the limit on effective neutrino self-coupling does not significantly improve with the addition of the 2015 Planck temperature and polarization data. As we will discuss further in section \ref{sec:discussion}, the strength of this limit is largely a function of the maximum CMB multipole included in the dataset. Since both Planck data releases extend to similar $l_{\rm max}$ ($\sim2500$), it is sensible that their constraining power of $\geff$ is similar.

Table \ref{table:results2} summarizes the parameter limits within the interacting neutrino mode. This mode is characterized by a higher value of the Hubble parameter $H_0\sim69.5$ km/s/Mpc, a lower value of the scalar spectral index $n_{\rm s}\sim0.935$, and a somewhat lower value of the scalar amplitude of fluctuation $10^9 A_{\rm s}\sim 2.15$. The mean value of $\geff$ within this mode can be translated to a neutrino decoupling redshift $z_{\nu,{\rm dec}}\sim8300$, that is, still significantly before matter-radiation equality. The values of $\Omega_{\rm b}h^2$, $\Omega_{\rm c}h^2$, and $\tau_{\rm reio}$ are broadly consistent with those of the $\Lambda$CDM mode. These results are similar to those presented in ref.~\cite{FYCR}. The main difference with this latter work is that the preferred value of $\log_{10}(\geff {\rm MeV}^2)$ we obtain is slightly higher than theirs, and the 1-sigma uncertainty on the mean has shrunk by approximately a factor of 2.  This is largely driven by the different calibration of the 2013 and 2015 Planck data \cite{planck15} and by the addition of CMB polarization data, as can be seen in the inset of figure \ref{fig:marggeff}. Given that the $E$-mode polarization data used in our analysis might contain systematics due to temperature-to-polarization leakage \cite{planck15}, we caution that this result should be interpreted with care. We expand on this point in section \ref{sec:discussion}. While CMB polarization data tend to raise the amplitude of the interacting neutrino mode, the addition of the BOSS DR12 BAO data \cite{alam16} tends to reduce its significance for all cases we considered. To understand why, note that the absence of a free-streaming phase shift \cite{Bashinsky:2003tk,Hou:2011ec,Follin:2015hya,Baumann15} for models in this corner of parameter space is compensated by adjusting the angular diameter distance to the last scattering surface (via a change in $H_0$). When a low redshift probe such as BAO is added, it provides an extra constraint in this angular-to-physical scale conversion, hence helping to break possible degeneracies. Our results suggest that improved low-redshift measurements of the BAO scale and of its phase \cite{Baumann:2017lmt} could help rule out the interacting neutrino mode as a viable possibility.

Interestingly, the higher value of the Hubble parameter preferred by the interacting neutrino mode helps alleviate the tension between CMB and local \cite{riess11,Riess:2016jrr} measurements of the present-day expansion rate. This is clearly illustrated in figure \ref{fig:triang_BAO_H0} where we show that the addition of the local Hubble parameter measurement to our analysis raises the statistical significance of the interacting neutrino mode. Comparing the values of $H_0$ given in table \ref{table:results2} with the latest measurement from ref.~\cite{Riess:2016jrr}, $H_0 =  73.24 \pm 1.74$ km s$^{-1}$ Mpc$^{-1}$, indicates that suppressing neutrino free-streaming reduces the tension but does not entirely remove it. Given that we have kept the effective number of neutrinos fixed to its standard value of $N_{\rm eff} = 3.046$ and have not included neutrino masses in our analysis, it is possible that nonstandard neutrino properties could play a key role in reconciling the CMB and local measurements of the Hubble parameter once the full parameter space of neutrino properties is explored. We leave such exploration to future work \cite{Kreisch_in_prep}. 

\subsection{Relative statistical weight of the two modes}\label{sec:bayesian_evidence}
We now turn our attention to the relative statistical importance of the interacting neutrino mode compared to the standard $\Lambda$CDM mode. For the purpose of this analysis, we define the interacting neutrino mode as occupying the range $\log_{10}\left( \geff {\rm MeV}^2\right) \in [-2.5, 0]$, while the $\Lambda$CDM mode consists of all models with $\log_{10}\left( \geff {\rm MeV}^2\right) < -2.5$. Upon comparison with figure \ref{fig:marggeff}, one can see that this is a reasonable splitting of the parameter space since the plane given by $\log_{10}\left( \geff {\rm MeV}^2\right) = -2.5$ corresponds to a  local minimum of the marginalized posterior. As a first statistical comparison, we compute the maximum-likelihood ratio $\rZ$ of the interacting neutrino mode to the $\Lambda$CDM mode, that is,
\begin{equation}
\rZ \equiv \frac{\max(\mathcal{L}_{\rm SI\nu})}{\max(\mathcal{L}_{\Lambda {\rm CDM}})},  
\end{equation}
where $\mathcal{L}_{\rm SI\nu}$ is the likelihood within the interacting neutrino mode, and $\mathcal{L}_{\Lambda {\rm CDM}}$ is the likelihood within the $\Lambda$CDM mode. We list the maximum-likelihood ratios for the different dataset combinations considered in this work in table \ref{tab:ratio}. We observe that the interacting neutrino mode is always significantly subdominant in terms of its maximum likelihood, except when both polarization data and the local $H_0$ measurement are included. In this case, it can reach $\sim82\%$ of the maximum likelihood of the $\Lambda$CDM model. This is largely due to the fact that the interacting neutrino mode favors higher values of $H_0$ which are more consistent with distance ladder measurement of $H_0$ from ref.~\citep{riess11}. It is also apparent from the ratios shown in table \ref{tab:ratio} that Planck polarization data tend to favor the interacting neutrino mode more strongly than the temperature data alone.  Additionally, as mentioned above, BAO data tend to suppress the maximum-likelihood ratio of the interacting neutrino mode compared to the standard $\Lambda$CDM mode.

\begin{deluxetable}{c|cccccc} 
\tabletypesize{\small} 
\tablecolumns{7} 
\tablewidth{0pt} 
\tablecaption{Maximum-likelihood ratio ($\rZ$) and Bayesian evidence ratio ($\Bfac$) of the interacting neutrino mode to the standard $\Lambda$CDM mode.  \label{tab:ratio}} 
\tablehead{ 
\colhead{$\mathcal{L}$} & \colhead{TT} &  \colhead{TT + Pol  } &  \colhead{TT + BAO  }& \colhead{TT + Pol} & \colhead{TT + BAO}  & \colhead{TT + Pol} \\
&&&&\colhead{+ BAO} &  \colhead{+ $H_0$} &   \colhead{+ BAO + $H_0$}} 
\startdata 
$\rZ$ & $0.27$ & $0.40$ & $0.10$ & $0.30$ & $0.24$ & $0.82$ \\
$\Bfac$&$0.10\pm0.01$ & $0.09\pm0.01$ & -      & $0.08\pm0.01$ & -      & $0.15\pm0.02$
\enddata 
\vspace{-0.7cm} 
\tablecomments{A $\rZ$ value of unity means that the interacting neutrino mode has a maximum data likelihood equivalent to the standard $\Lambda$CDM model. A $\Bfac$ value less than unity implies that the $\Lambda$CDM model is favored over the interacting neutrino cosmology. Due to the significant computational burden, $\Bfac$ was computed only for certain data combinations considered.} 
\end{deluxetable}

A perhaps more robust measure of the relative statistical importance of the interacting neutrino mode can be obtained by computing its Bayesian evidence, and comparing it to that of the $\Lambda$CDM mode. The Bayesian evidence is defined in terms of an integral over the entirety of the parameter space as
\begin{equation}
\label{evidence}
\mathcal{Z} \equiv \int_{\Omega_{\mathcal{M}}} \mathcal{L}(d|\boldsymbol{\theta},\mathcal{M}) \Pi(\boldsymbol{\theta}|\mathcal{M}) \diff \boldsymbol{\theta},
\end{equation}
where $\boldsymbol{\theta}$ is the vector of parameters, $\Omega_{\mathcal{M}}$ is the volume of the model's parameter space, $\mathcal{L}(d|\boldsymbol{\theta},\mathcal{M})$ is the probability of the data given the parameters and the model, or the likelihood function, and $\Pi(\boldsymbol{\theta}|\mathcal{M})$ is the prior distribution placed on the model.  Note that the Bayesian evidence is then just as dependent on the priors as it is on the model, $\mathcal{M}$.  In this way we can define a Bayesian evidence for the $\Lambda$CDM mode, as well as separately for the interacting neutrino mode based on the choice of $\log_{10}\left( \geff {\rm MeV}^2\right)$ prior given at the beginning of this subsection. We perform separate MultiNest \cite{feroz08,feroz09,feroz13} runs for each mode of the posterior, requesting a $10\%$ tolerance on the Bayesian evidence estimate provided by the code. These results are then used to compute the Bayes factor\footnote{Note that we are not using the Bayes factor for model comparison (as it is commonly used), but rather to compare the statistical weight of two modes of the same model.}, which is defined as
\begin{equation}
\label{bayesfactor}
\Bfac \equiv \frac{\mathcal{Z}_{\rm SI\nu}}{\mathcal{Z}_{\Lambda{\rm CDM}}},
\end{equation}
where $\mathcal{Z}_i$ refers to the Bayesian evidence of model $i$ according to the definition in eq.~\eqref{evidence}. Essentially, $\Bfac$ is the ratio of the average likelihood within each mode. We list in table \ref{tab:ratio} the values of $\Bfac$ and their uncertainty for the different dataset combinations considered in this work. In all cases, the Bayes factor is much less than unity, implying that the data favor the standard $\Lambda$CDM mode over the interacting neutrino cosmology. 

We thus conclude, assuming massless neutrinos with $N_{\rm eff} = 3.046$ and a flat prior on $\log_{10}\left( \geff {\rm MeV}^2\right)$, that current data show no statistical evidence for the presence of strong self-interaction in the neutrino sector. While it is intriguing that the likelihoods within the two distinct modes are somewhat comparable (especially when local $H_0$ measurements are included), the small volume of parameter space where the interacting neutrino mode provides a good fit to the data ultimately results in a Bayes factor favoring the $\Lambda$CDM mode. We note that the maximum-likelihood ratio and the Bayes factor show slightly different trends when moving from one dataset to the next, but our conclusions are unaffected by this scatter as the numbers remain small nonetheless. While we do not consider here the impact of an alternate choice of priors, it is clear that using, for instance, a flat prior on $\geff$ itself could dramatically alter our conclusions. Arguably, our current choice of priors could be considered conservative since it does not pick a particular energy scale for the new neutrino physics. As long as no external information is known about new physics in the neutrino sector, we strongly believe that a flat, non-informative prior on $\log_{10}\left( \geff {\rm MeV}^2\right)$ is the most reasonable choice. Of course, it would be interesting to determine how the statistical significance of the interacting neutrino mode is affected once neutrino masses are included and $N_{\rm eff}$ is allowed to deviate from its standard value.
%
\section{Discussion}\label{sec:discussion}
%
\subsection{Bimodality}

\begin{figure}[t!]
\begin{center}
\includegraphics[width=0.8\textwidth]{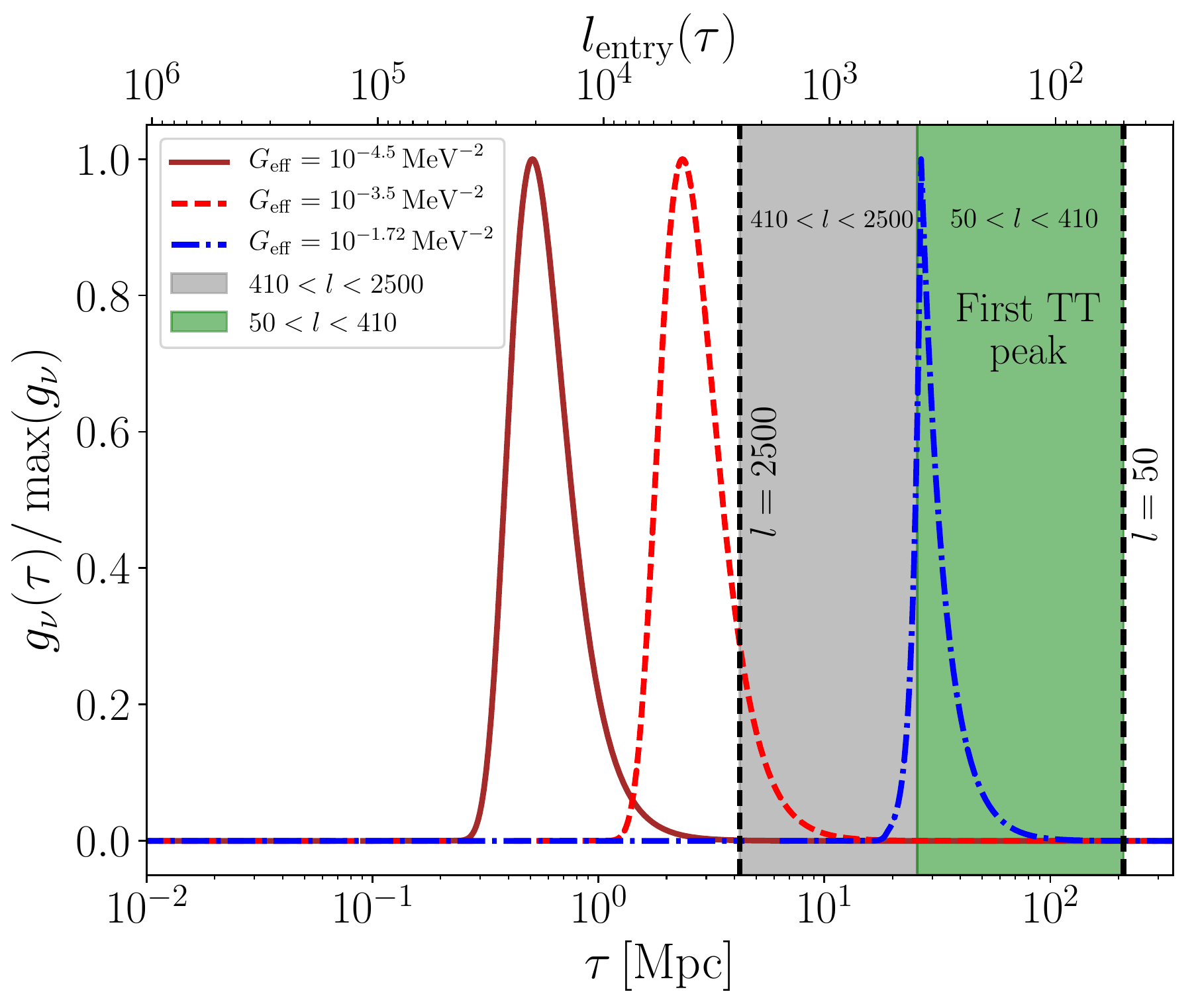}
\caption{Neutrino visibility function $g_\nu(\tau) = -\dot{\tau}_\nu e^{-\tau_\nu}$ as a function of conformal time $\tau$. The top $x$-axis shows the approximate CMB multipole $l_{\rm entry}$ that is entering the causal horizon at each value of $\tau$ according to the relation $l_{\rm entry} = 0.75(\tau_0/\tau)$ \cite{Dodelson-Cosmology-2003}, where $\tau_0$ is the conformal time today. The solid brown line shows a model where neutrino free streaming occurs early enough as to not affect the CMB, while the red dashed line displays the model which corresponds to the 95\% upper bound on $\geff$ from the $\Lambda$CDM mode. The dot-dashed blue line shows the neutrino visibility function best fit from the interacting neutrino mode. The grayed region shows the approximate range of conformal time probed by the the CMB multipole $410<l<2500$, while the green region illustrates the range probed by the full width of the first CMB temperature peak ($50<l<410$).}
\label{fig:vis_funct}
\end{center}
\end{figure}

One of the key questions that our results raise is why we obtain a bimodal distribution. On the one hand, the existence of a mode consistent with the standard $\Lambda$CDM cosmology in which neutrinos begin free-streaming before the onset of Big Bang nucleosynthesis is not too surprising. That this mode spans values of the effective neutrino coupling constant that are more than seven orders of magnitude above the standard Fermi constant is simply a reflection that the scales probed by the current CMB data are insensitive to the onset of neutrino free-streaming if it happens early enough. Indeed, for $\geff \lesssim 10^{-4.5}$MeV$^{-2}$, neutrino decoupling occurs before the Fourier modes probed by the Planck data enter the causal horizon, implying that they are unaffected by the new neutrino interactions and receive the standard phase and amplitude shift associated with neutrino free-streaming. We illustrate this in figure \ref{fig:vis_funct} where we show the neutrino visibility function $g_\nu(\tau) \equiv -\dot{\tau}_\nu e^{-\tau_\nu}$ as a function of conformal time.\footnote{Much like the better-known CMB visibility function, the neutrino visibility function is a probability density function for the time at which neutrinos last scatter.} The gray band shows the approximate time interval in which the multipoles $410<l<2500$ enter the causal horizon. This multipole range corresponds to scales encompassing all well-measured CMB temperature peaks except for the first one. We see that the visibility functions of models with $\geff \lesssim 10^{-4.5}$MeV$^{-2}$ have no overlap with the time interval at which the modes probed by the current Planck data are entering the horizon. This explains why the posterior shown in figure \ref{fig:marggeff} flattens out for this range of neutrino self-interacting strength.

The sharp suppression of the $\Lambda$CDM mode of the posterior distribution for $\geff > 10^{-4.5}$MeV$^{-2}$ indicate that these values delay neutrino free-streaming long enough for the length scales probed by the CMB damping tail to enter the horizon. This is supported by figure \ref{fig:vis_funct} where the red dashed line shows the models corresponding to the 95\% upper limit of the $\Lambda$CDM mode ($\geff\approx 10^{-3.5}$MeV$^{-2}$), whose visibility function has significant overlap with the modes probed by the CMB. This indicates that the allowed upper limit on $\geff$ within the $\Lambda$CDM mode strongly depends on the highest multipole probed by the data since  higher $l_{\rm max}$ are capable of probing an earlier onset of neutrino free streaming and thus smaller values of $\geff$. It is thus not surprising that our constraint on the $\Lambda$CDM mode is similar to that from ref.~\cite{FYCR} since the value of $l_{\rm max}$ between the Planck 2013 and 2015 data release did not appreciably change.

\begin{figure}[t!]
\begin{center}
\includegraphics[width=0.495\textwidth]{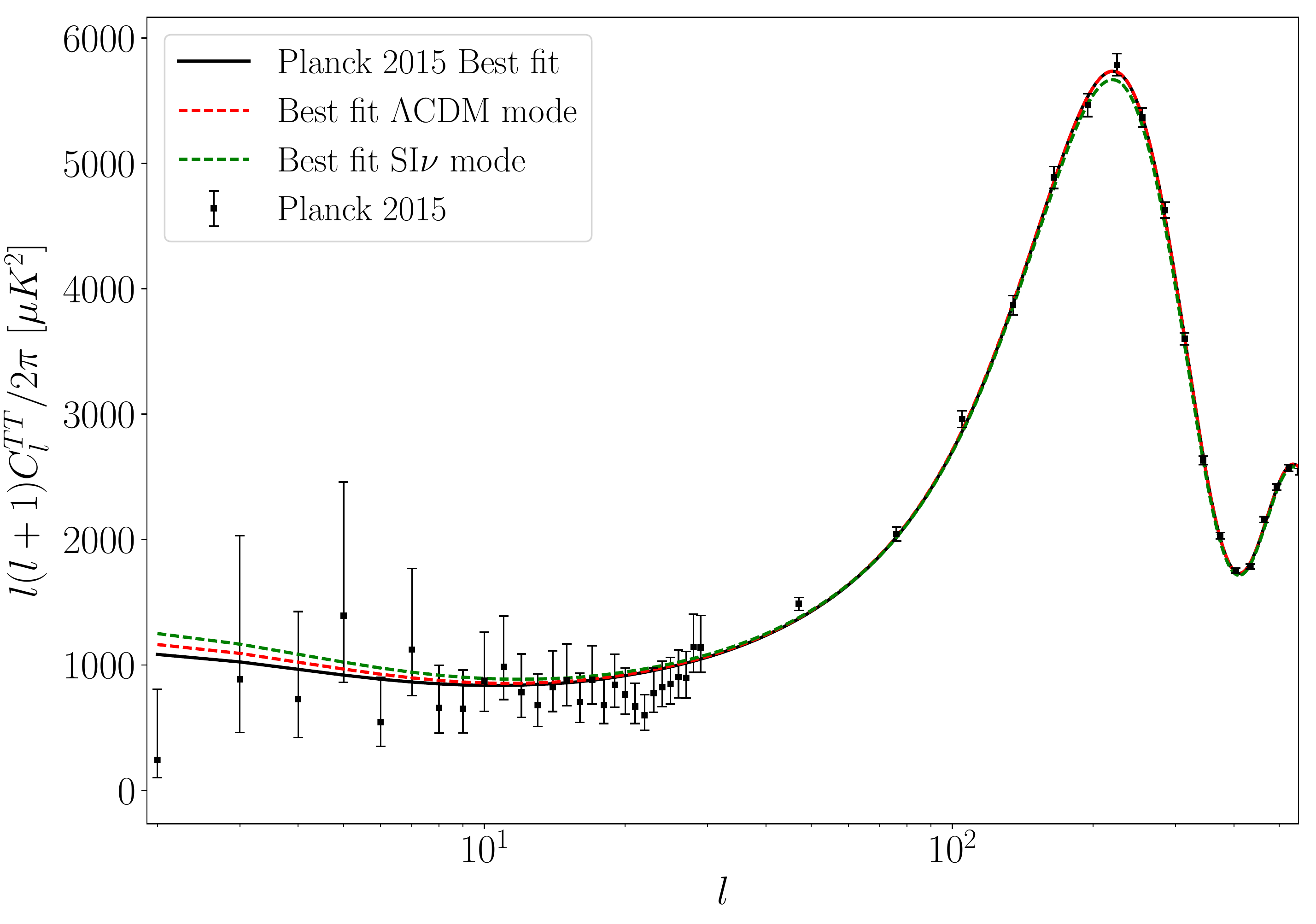}
\includegraphics[width=0.48\textwidth]{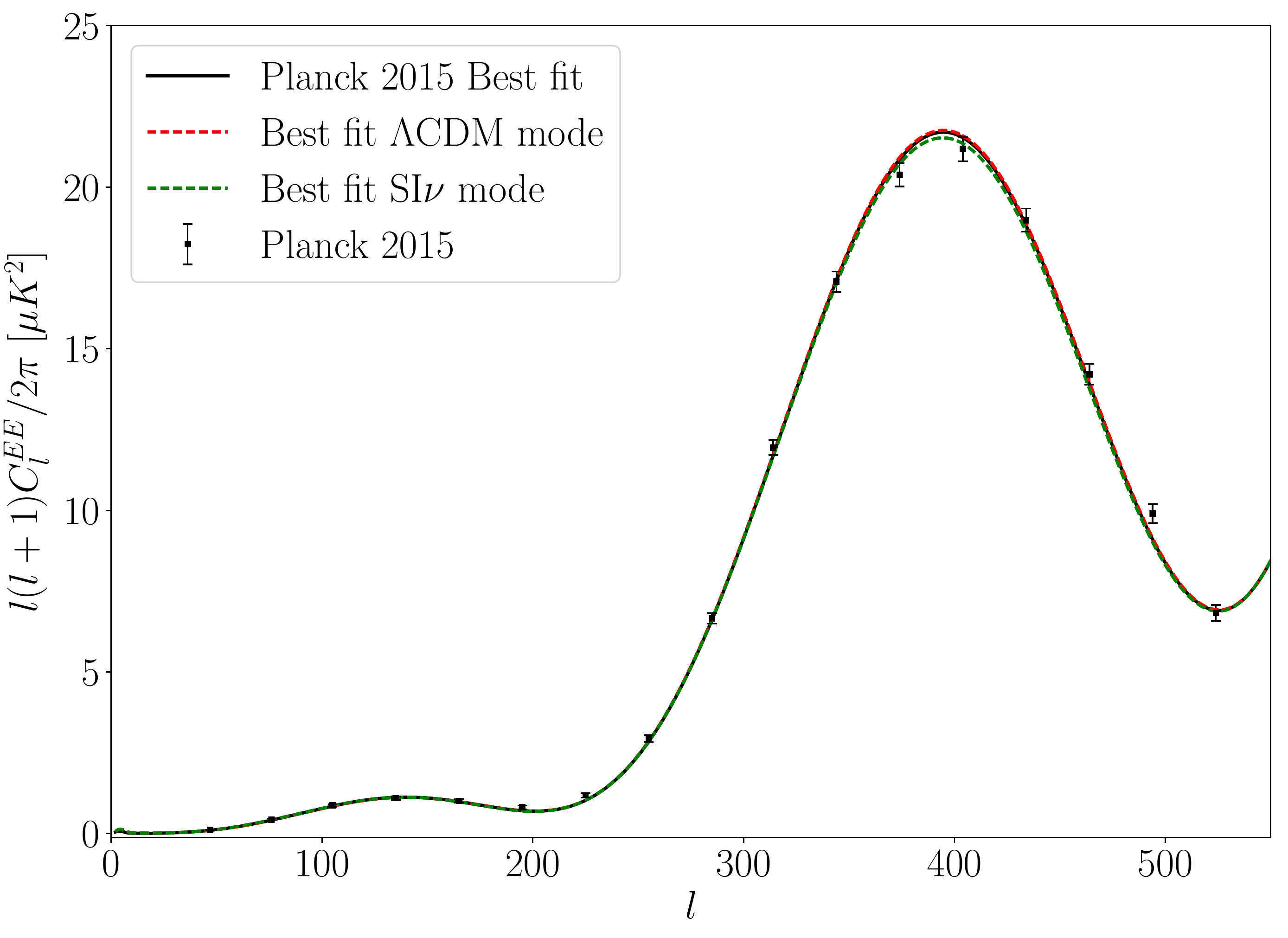}
\caption{Low-$l$ CMB temperature (left panel) and $E$-mode polarization (right panel) power spectra as a function of angular multipole. The black points are the Planck 2015 data \cite{planck15Params}. The red dashed line shows the best-fit model from the $\Lambda$CDM mode using the data combination ``TT + Pol + BAO + $H_0$'', while the dashed green line displays the best fit model within the interacting neutrino (SI$\nu$) mode for the same choice of data.}
\label{fig:TTEEPowSpec_lowl}
\end{center}
\end{figure}
\begin{figure}[t!]
\includegraphics[width=\textwidth]{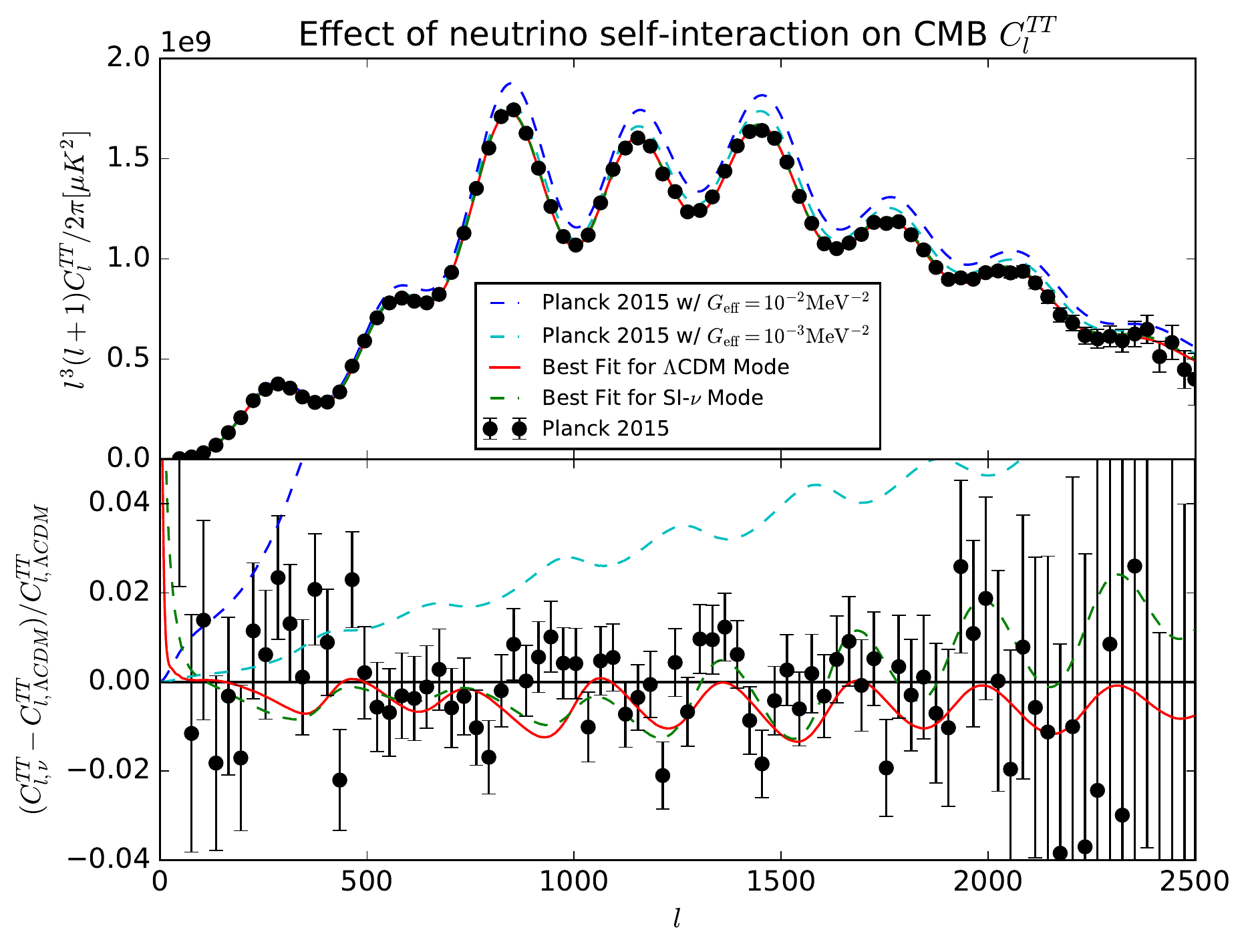}
\caption{CMB temperature power spectrum as a function of angular multipole. The black points are the Planck 2015 data \cite{planck15Params}. The red line shows the best-fit model from the $\Lambda$CDM mode using the data combination ``TT + Pol + BAO + $H_0$'', while the dashed green line displays the best fit model within the interacting neutrino (SI$-\nu$) mode for the same choice of data. For comparison, we also show models with $\geff = 10^{-2} $ MeV$^{-2}$ and $\geff = 10^{-3} $ MeV$^{-2}$ (dashed blue and cyan lines, respectively) for which all other cosmological parameters are set according to the Planck 2015 best fit $\Lambda$CDM model using their data combination ``TT, TE, and EE + lowP'' as described in table 3 of ref.~\cite{planck15}. The bottom panel shows the fractional residuals of these four models with respect to this best fit $\Lambda$CDM cosmology from Planck 2015.}
\label{fig:TTPowSpec}
\end{figure}

The deep trough in the posterior distribution for $-3.2 \lesssim \log_{10} \left( \geff {\rm MeV}^2\right)\lesssim-2.3$ indicate that CMB data strongly disfavor neutrino decoupling occurring while the modes corresponding to the Silk damping tail are entering the horizon. Indeed, models with $\geff$ in this range have a neutrino visibility function peaking within the gray band of figure \ref{fig:vis_funct}. The dot-dashed blue line shows the neutrino visibility function for the best fit model within the  interacting neutrino mode. We observe that this visibility function peaks right as the multipoles corresponding to the first CMB temperature peak (green shaded region) begin to enter the Hubble horizon. In this case, none of the CMB temperature peaks in the range $410<l<2500$ receives the phase and amplitude shift usually associated with neutrino free streaming, hence requiring the other cosmological parameters, notably $H_0$, $A_{\rm s}$, and $n_{\rm s}$, to absorb the resulting difference in the temperature spectrum (see figure \ref{fig:TTPowSpec}). From the perspective of CMB polarization, the visibility function of the best fit interacting neutrino model has a maximum near the epoch when the second peak of the $E$-mode polarization spectrum at $l\approx 370$ is entering the horizon. 

To understand the impact of this late neutrino decoupling, it is instructive to look at the CMB temperature and $E$-mode polarization spectra on the relevant scales ($l\lesssim410$), as shown in figure \ref{fig:TTEEPowSpec_lowl}. There, we observe that the interacting neutrino mode predicts a slightly lower amplitude for the first peak of the temperature spectrum compared to the standard $\Lambda$CDM cosmology, while displaying more power than the standard paradigm at low multipoles. This indicates that the degeneracy with the cosmological parameters $H_0$, $A_{\rm s}$, and $n_{\rm s}$ is not exact and that the interacting neutrino mode is a compromise between having a large enough amplitude of the first temperature peak while not overproducing power at very low multipoles. This actually hints at why the interacting neutrino mode is statistically subdominant compared to the $\Lambda$CDM mode: it leads to a worse fit of the low-$l$ temperature data, hence suppressing its overall likelihood. This also indicates why the interacting neutrino mode is only viable for such a small range of $\geff$: larger values of $\geff$ require even lower values of $n_{\rm s}$ and $A_{\rm s}$ and even higher values of $H_0$ in order to fit the $C_l^{TT}$ damping tail and first peak, which invariably leads to a CMB temperature spectrum with too much power at low multipoles. 

The picture that emerges is that self-interacting neutrinos are only viable if they either decouple before the bulk of the modes probed by the CMB enter the horizon, or if they decouple within a narrow window centered around $l\sim400$. \emph{In all cases, neutrinos must begin to free stream before matter-radiation equality} (even within the interacting neutrino mode, $z_{\nu,{\rm dec}}\approx 8300$). For this latter case, the large cosmic variance-dominated error bars at low multipoles allow for an approximate multi-parameter degeneracy to correct for the impact of self-interacting neutrino on the CMB damping tail, at the price of having a slightly lower amplitude of the first temperature peak. It is however remarkable that the parameter changes required to make the interacting neutrino mode fit the Planck temperature data reasonably well also lead to an $E$-mode polarization spectra that is in very good agreement with the data. This is illustrated in the right panel of figure \ref{fig:TTEEPowSpec_lowl}  (see also next subsection). 

\subsection{Broad structure of temperature and polarization spectra}
\begin{figure}[t!]
\includegraphics[width=1.00\textwidth]{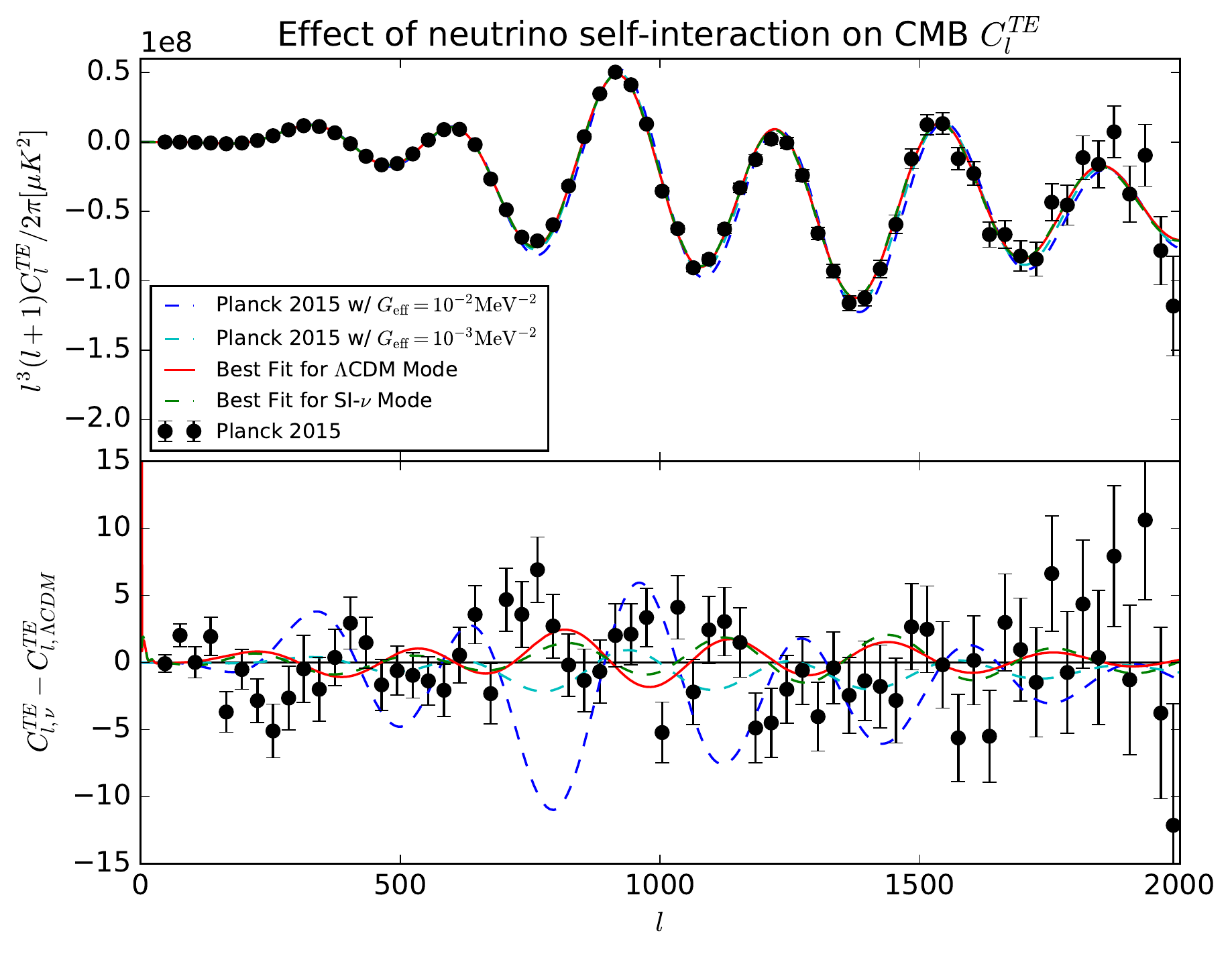}
\caption{Similar to figure \ref{fig:TTPowSpec} but this time showing the cross-power spectrum of temperature and $E$-mode polarization. Note that the bottom panel now shows absolute residuals between the different interacting neutrino models and the best fit Planck $\Lambda$CDM cosmology.}
\label{fig:TEPowSpec}
\end{figure}
\begin{figure}[t!]
\includegraphics[width=1.00\textwidth]{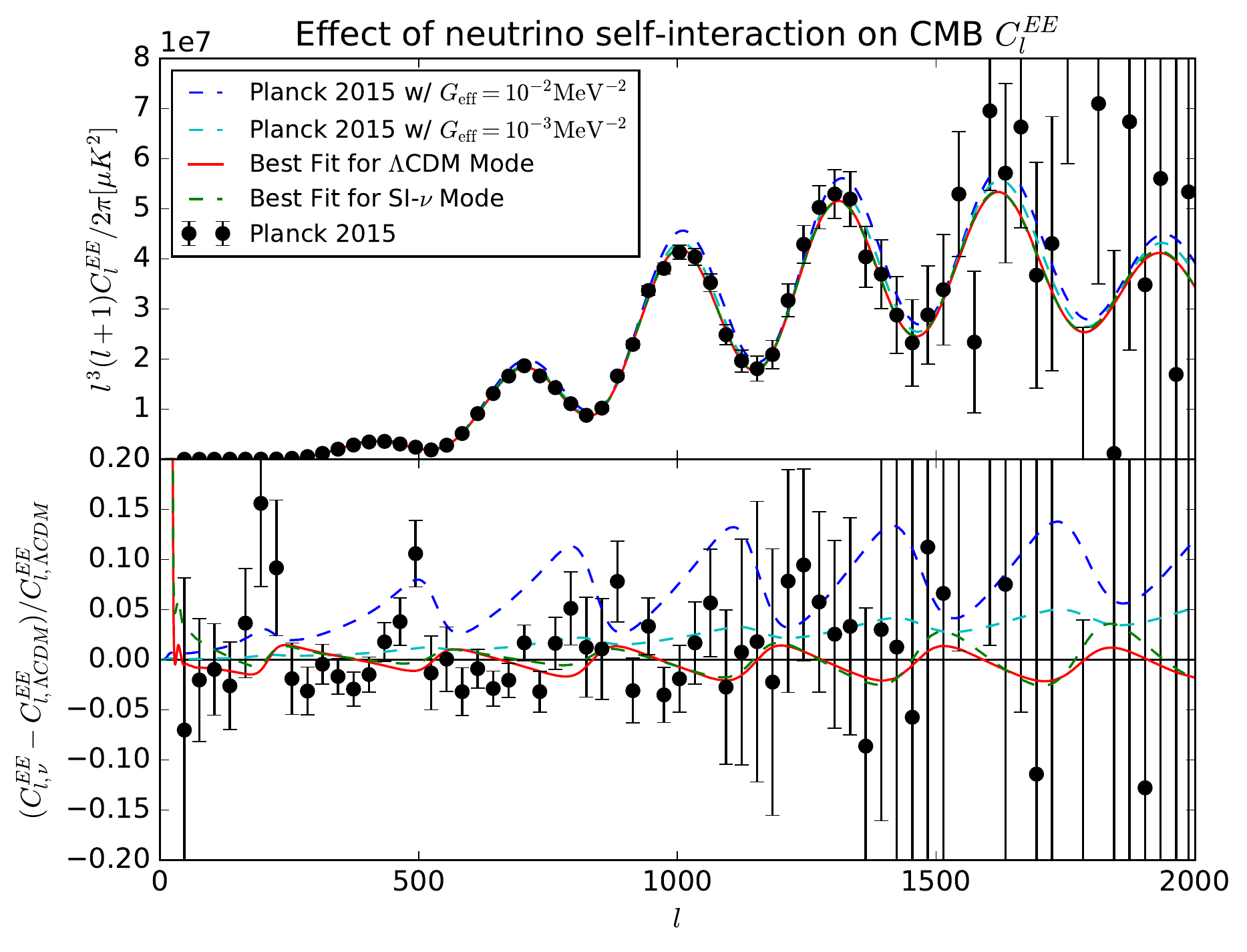}
\caption{Similar to figure \ref{fig:TTPowSpec} but now showing the $E$-mode polarization power spectrum.}
\label{fig:EEPowSpec}
\end{figure}

We compare in figure \ref{fig:TTPowSpec} the CMB temperature anisotropy spectrum of the best fit models for both the $\Lambda$CDM mode and the interacting neutrino mode, along with two reference models with $\geff = 10^{-2}$MeV$^{-2}$ and $\geff = 10^{-3}$MeV$^{-2}$ for which the other cosmological parameters are kept fixed at their Planck 2015 values. The lower panel illustrates the fractional residuals between these four models and the Planck 2015 \cite{planck15} best fit model from their ``TT, TE, EE + lowP'' data combination. The first thing to notice is that the residuals of the best fit models of the two modes show a similar oscillatory pattern, but with the interacting neutrino mode displaying more power at large multipoles, where the error bars are relatively big. Over the  range $50\leq l\leq 2000$, the best fit interacting neutrino cosmology does not deviate from the Planck 2015 model by more than 2\%, which is somewhat remarkable given how radically different this model is compared to the standard $\Lambda$CDM paradigm. This significant difference can be understood by examining the two other plotted models ($\geff = 10^{-2}$MeV$^{-2}$ and $\geff = 10^{-3}$MeV$^{-2}$) for which the large amplitude and phase shift are readily noticeable. The existence of a degeneracy between the neutrino self-interaction strength and the parameter combination $A_{\rm s}$, $n_{\rm s}$, and $H_0$ that nearly compensates for these large changes to the CMB temperature spectrum is remarkable. It also highlights the importance of correctly modeling neutrino physics when performing CMB analyses since this degeneracy is missed when the popular $\{c_{\rm eff},c_{\rm vis}\}$ parametrization is used to study deviation from standard free-streaming neutrinos. 

Figures \ref{fig:TEPowSpec} and \ref{fig:EEPowSpec} display the $C_l^{TE}$ and $C_l^{EE}$ CMB spectra, respectively, for the same four models. Again, we observe that the best fit model within the interacting neutrino mode (green dashed line) has residuals similar to that of the best fit model within the $\Lambda$CDM mode (red solid line). This indicates that the multi-parameter degeneracy responsible for compensating the impact of neutrino self-interaction on the CMB temperature spectrum also leads to an $E$-mode polarization spectrum that closely matches the Planck data. As noted in table \ref{table:results2}, including polarization data does yield a slightly higher value of the neutrino self-interaction strength, but this shift is entirely consistent with the value inferred from temperature-only data at less than 1-$\sigma$. We caution though that the presence of potential systematics within the Planck polarization data could affect our results, and they should therefore be interpreted with care. However, since our results with and without the Planck polarization data are largely consistent with one another, it is very unlikely that these systematics would dramatically change our conclusions. Indeed, we emphasize that our upper limit on $\log_{10} \left( \geff {\rm MeV}^2\right)$ within the $\Lambda$CDM mode and the existence of the somewhat subdominant interacting neutrino mode are both robust to the possible presence of systematics in the data. 

\section{CMB Stage-IV forecast}\label{forecasting}
We end our analysis by performing a Fisher matrix forecast for the expected sensitivity of Stage-IV CMB experiments (CMB-S4) \cite{2016arXiv161002743A}. The Fisher formalism assumes a Gaussian PDF for the parameters. Since our posterior is bimodal and thus cannot be accurately approximated by a single multivariate Gaussian, we perform separate forecasts for each mode of the posterior. We emphasize that our Fisher analysis cannot be used to determine whether future data can rule out the interacting neutrino mode. Instead, it is used to provide an estimate of how much future data will tighten the parameter constraints within each mode. Following ref.~\citep{wufisher14}, we assume a Gaussian likelihood $\mathcal{L}$ for the parameter vector
$\boldsymbol{\theta}$ given some data vector $\mathbf{d}$
\begin{equation}
\label{likelihood}
\mathcal{L}(\boldsymbol{\theta}|\mathbf{d}) \propto \frac{1}{\sqrt{\det \mathbf{C}(\boldsymbol{\theta}) }}\exp \left( - \frac{1}{2} \mathbf{d}^{\dagger}\left[\mathbf{C}(\boldsymbol{\theta})\right]^{-1} \mathbf{d}\right),
\end{equation}
where $\mathbf{C}(\boldsymbol{\theta})$ is the covariance matrix of the data vector.
The Fisher matrix itself is then built by taking second-order partial derivatives of the likelihood,
or its curvature, at the fiducial values of the parameters $\boldsymbol{\theta}_0$
\begin{equation}
\label{fisher1}
F_{ij} = \left\langle \left. \frac{\partial^2 \log \mathcal{L}}{\partial \theta_i \partial \theta_j} \right|_{\boldsymbol{\theta}=\boldsymbol{\theta}_0} \right\rangle ,
\end{equation}
and the uncertainty of parameter $\theta_i$ is given by
\begin{equation}
\sigma_i \equiv \sigma(\theta_i) = \sqrt{(\mathbf{F}^{-1})_{ii}}.
\end{equation}
For CMB experiments, we choose our data vector to be $\mathbf d= \{ a^T_{lm},a^E_{lm}\}$,
with $a^T_{lm},a^E_{lm}$ being
the spherical harmonic coefficients of the temperature field and the E-mode polarization field, respectively;
consequently its covariance matrix for each multipole is given by
\[\mathbf{C}_l(\boldsymbol{\theta}) \equiv \left(\begin{array}{ccc}
C_l^{TT} + N_l^{TT} & C_l^{TE} \\
C_l^{TE} & C_l^{EE} + N_l^{EE}
\end{array}
\right),\]
where the Gaussian noise $N_l^{XX}$ is defined as
\begin{equation}
N_l^{XX} = \Delta_X^2 \exp \left( l(l+1) \frac{\theta^2_{\rm FWHM}}{8\log 2}\right),
\end{equation}
with $\Delta_X (X = T, E)$ being the experiment pixel noise level,
and $\theta_{\rm FWHM}$ being the full width at half-maximum beam size in radians.
Combining the above eqns.~\eqref{likelihood} and \eqref{fisher1},
we find
\begin{equation}
F_{ij} = \sum_{l = l_{\rm min}}^{l_{\rm max}} \frac{2l + 1}{2} f_{\rm sky} {\rm Tr}
\left( \mathbf{C}_l^{-1}(\boldsymbol{\theta})
\frac{\partial \mathbf{C}_l}{\partial \theta_i}\mathbf{C}_l^{-1}(\boldsymbol{\theta})
\frac{\partial \mathbf{C}_l}{\partial \theta_j} \right),
\end{equation}
where $f_{\rm sky}$ is the fraction of covered sky.
In our forecast, we take partial derivatives directly over $G_{\rm eff}$ rather than its logarithm as this is better suited to the linear approximation of the Fisher formalism. The fiducial parameters $\boldsymbol{\theta}_0$ for each mode are chosen as the mean values from our inference analysis using the `` TT + Pol'' data combination (see tables \ref{table:results1} and \ref{table:results2}). The only exception is the value of $\geff$ in the $\Lambda$CDM mode, which we take to be $\geff = 10^{-4.5}$ MeV$^{-2}$. We assume the following specifications for CMB-S4: $f_{\rm sky} = 0.5$, $30\leq l\leq3000$, $\Delta_T = 1.5 \mu$K-arcmin, $\Delta_E = \sqrt{2}\Delta_T$, and $\theta_{\rm FWHM} = 1$ arcmin. We list the forecasted 1-$\sigma$ error bars for each mode in table \ref{tab:fisher}. Our forecast suggests that CMB-S4 could significantly improve the upper bound on $\geff$ within the $\Lambda$CDM mode, with a projected limit of $\log_{10}(\geff{\rm MeV}^2) < -4.0$ at 95\% confidence level. We note that this improvement is largely driven by the larger $l_{\rm max}$ and the smaller error bars of the projected CMB-S4 dataset.
\begin{deluxetable}{c|cc|cc} 
\tabletypesize{\small} 
\tablecolumns{5} 
\tablewidth{0pt} 
\tablecaption{ CMB-S4 forecast for the 1-$\sigma$ error bars for both the $\Lambda$CDM and interacting neutrino modes.\label{tab:fisher}} 
\tablehead{ 
& \multicolumn{2}{c}{$\Lambda$CDM mode } &  \multicolumn{2}{c}{Interacting neutrino mode}\\
\colhead{Parameters} &\colhead{\emph{Fiducial values}} &  \colhead{\emph{1-$\sigma$ CMB-S4}} &  \colhead{\emph{Fiducial values}} &  \colhead{\emph{1-$\sigma$ CMB-S4}} } 
\startdata 
$\Omega_{\rm b} h^2$ & $0.02223$ & $0.00003 $ & $0.02248$& $0.00003$ \\ 
$\Omega_{\rm c} h^2$ & $0.1193$ & $0.0004$& $0.1200 $  & $0.0003$ \\ 
$H_0$ [km/s/Mpc] & $67.9 $ & $0.2$  & $69.6 $ & $0.2$ \\ 
$\tau$  & $0.095 $ & $ 0.003$ &  $0.103 $ & $0.003$\\ 
$n_{\rm s}$ & $0.962 $ & $0.001$ &$0.938 $  & $0.001$ \\ 
$10^9A_{\rm s}$ & $2.27 $& $0.01  $ & $2.16 $  & $0.01 $\\
$10^2 G_{\rm eff}\ [{\rm MeV}^{-2}]$ & $ 0.0032$ & $0.0033$ & $1.87 $ & $0.07$\\ 
\enddata 
\vspace{-0.4cm} 
\end{deluxetable}
%

\section{Conclusion}\label{sec:conclusion}

In this paper, we have presented updated cosmological bounds on the free-streaming nature of neutrinos in the early Universe, focusing on models where neutrino self-scattering is well-described by a Fermi-like four-fermion interaction. Using the Planck 2015 CMB temperature and polarization data, the BAO measurement from BOSS DR12, and the local measurement of a Hubble parameter, we have determined that the onset of neutrino free streaming must occur either before any of the Fourier modes probed by the CMB enter the causal horizon ($z_{\nu,{\rm dec}}\gtrsim1.3\times10^5$) or within a narrow window centered around redshift $z_{\nu,{\rm dec}}\sim 8300$. This latter case (i.e.~the interacting neutrino mode) is possible due to a multi-parameter degeneracy between the neutrino self-interaction strength and the combination of parameters $\{A_{\rm s},n_{\rm s},H_0\}$. This degeneracy is however not exact, and the interacting neutrino mode ends up predicting an excess of power in the temperature spectrum at low multipoles compared to what is observed by Planck, leading to the mode being statistically subdominant compared to the more standard $\Lambda$CDM mode. In addition, the BAO measurement considered in this work also tends to disfavor the interacting neutrino mode, while the Planck $E$-mode polarization data has the opposite effect and increase the statistical significance of this secondary mode. Interestingly, the interacting neutrino cosmology can naturally accommodate a higher value of the Hubble parameter, and can thus help reconcile to some extent the tension between the local measurements based on a distance ladder with those derived from CMB data.

The surest way to improve on the constraints presented here is to extend the measurement of the CMB damping tail to higher value of $l_{\rm max}$. While such measurements already exist (e.g.~\cite{Story:2012wx,Das:2013zf}) and are bound to improve in the near future, the presence of significant foregrounds at $l\gtrsim2500$ in the temperature spectrum might partially nullify the gain of going to smaller scales. However, it is likely that better CMB polarization measurements on small scales could significantly improve the present constraints, as shown with our Fisher forecast for CMB-S4. In any case, probing larger multipoles within the CMB damping tail and reducing the measurement error bars could help rule out the interacting neutrino mode since the latter tends to display more power on small scales compared to its $\Lambda$CDM counterpart (see figures \ref{fig:TTPowSpec} and \ref{fig:EEPowSpec}). Including additional low-redshift probes could also in principle help further constraining the interacting neutrino mode. We note however that care must be taken when using these types of probes since the late neutrino decoupling within the interacting neutrino mode modifies the matter power spectrum \cite{Kreisch_in_prep} on scales where BAO, galaxy clustering, and weak lensing all have significant sensitivity. Since the analyses performed on the raw data often assumed a $\Lambda$CDM background cosmology, (see e.g.~ref.~\cite{alam16}) the constraints on derived parameters such as the BAO scale or $\sigma_8$ extracted from these analyses are difficult to interpret within the interacting neutrino mode context. On the bright side, the continued detection of ultra-high energy neutrinos at IceCube could eventually ruled out this peculiar mode of the $\geff$ posterior while significantly improving the bounds within the $\Lambda$CDM mode \cite{Ng:2014pca,Cherry:2016jol}. 

In this work, we have focused for simplicity on massless neutrinos with a standard thermal history (that is, we fixed $N_{\rm eff} = 3.046$). In reality, we know that neutrinos have mass and that the introduction of non-standard neutrino interaction is likely to change their thermal history and lead to a different value for the effective number of neutrinos.  On the one hand, we expect our constraints on the largest possible value of $\geff$ within the $\Lambda$CDM mode to be robust to the introduction of neutrino mass since neutrinos are still ultra-relativistic at the onset of free streaming in this case. On the other hand, we expect neutrino masses to have a larger effect on the interacting neutrino mode since neutrino decoupling within this model occurs close to the epoch at which neutrinos become nonrelativistic. It would be interesting to see how the statistical significance of the interacting neutrino mode changes once the neutrino mass and $N_{\rm eff}$ are allowed to vary, especially when local Hubble parameter measurements are included. We leave such consideration to future work \cite{Kreisch_in_prep}.

\noindent \textbf{Note added at Publication:} We would like to note that after the submission of this work, the preprint \cite{Oldengott17} appeared on arXiv, which performs a very similar analysis to this work. The main differences between \cite{Oldengott17} and this work are their exact treatment of the details of the momentum dependence of the neutrino interaction under study here and how varying $N_{\rm eff}$ changes the derived results.  While this work employs a much more detailed analysis of this interaction than our work, the results that are obtained are in good agreement with ours.  As is pointed out in \cite{Oldengott17}, this provides a good justification for the thermal approximation of the collision term of the Boltzmann hierarchy that we employ here.

\section*{Acknowledgements}
The work of this paper could not have been completed without the advice and review provided by members of the Knox research group: Dr.~Marius Millea and Brigid Mulroe. We would especially like to thank Dr.~Joe Zuntz, leader of the CosmoSIS team, for his excellent and quick feedback on technical issues. We thank Shadab Alam for his help putting together the likelihood updates for the most recent BOSS DR12 data. We also thank the McWilliams Center for Cosmology, for the use of their `Coma Cluster' computer cluster. F.-Y. C.-R. wishes to thank Prateek Agrawal for useful discussions. The work of L.~L. was supported in part by NSF grant PHY-1263201 as part of the Research Experience for Undergraduates (REU) program. L. L.wishes to thank the University of California, Davis, for hospitality while part of this work was completed. F.-Y. C.-R.~acknowledges the support of the National Aeronautical and Space Administration ATP grant NNX16AI12G at Harvard University.

\bibliographystyle{JHEP}
\bibliography{Interacting_neutrinos}

\end{document}